\pdfoutput=1

\documentclass[11pt,nofootinbib]{article}
\usepackage{amsmath,amssymb,mathtools}
\usepackage{graphicx}
\usepackage[colorlinks=true,linkcolor=blue,citecolor=blue,urlcolor=blue,linktocpage=true]{hyperref}
\usepackage{subfigure}
\usepackage{comment}
\usepackage{tensor}
\usepackage{cases}

\numberwithin{equation}{section}

\begin{document}

\newcommand{\comm}[1]{\textcolor{red}{\textbf{#1}}}

\providecommand{\abs}[1]{\lvert#1\rvert}
\providecommand{\bd}[1]{\boldsymbol{#1}}

\begin{titlepage}

\setcounter{page}{1} \baselineskip=15.5pt \thispagestyle{empty}

\begin{flushright}
SISSA 33/2017/FISI
\end{flushright}
\vfil

\bigskip
\begin{center}
 {\LARGE \textbf{Lyman-$\alpha$ Constraints on Ultralight Scalar Dark Matter:}}\\
\medskip 
 {\LARGE \textbf{Implications for the Early and Late Universe}}
\vskip 15pt
\end{center}

\vspace{0.5cm}
\begin{center}
{\Large 
Takeshi Kobayashi,$^{\star, \dagger}$
Riccardo Murgia,$^{\star, \dagger}$
Andrea De Simone,$^{\star, \dagger}$
Vid Ir\v{s}i\v{c},$^{\ast}$
and
Matteo Viel$^{\star, \dagger, \ddagger}$
}\end{center}

\vspace{0.3cm}

\begin{center}
\textit{$^{\star}$ SISSA, Via Bonomea 265, 34136 Trieste, Italy}\\

\vskip 14pt
\textit{$^{\dagger}$ INFN, Sezione di Trieste, Via Bonomea 265,
 34136 Trieste, Italy}\\ 

\vskip 14pt
 \textit{$^{\ast}$ University of Washington, Department of Astronomy,\\
 3910 15th Ave NE, WA 98195-1580 Seattle, USA}\\
 
\vskip 14pt
 \textit{$^{\ddagger}$ INAF/OATS, Osservatorio Astronomico di Trieste, via Tiepolo 11, I-34143 Trieste, Italy}\\   

\vskip 14pt
E-mail:
 \texttt{\href{mailto:takeshi.kobayashi@sissa.it}{takeshi.kobayashi@sissa.it}},
 \texttt{\href{mailto:riccardo.murgia@sissa.it}{riccardo.murgia@sissa.it}},
 \texttt{\href{mailto:andrea.desimone@sissa.it}{andrea.desimone@sissa.it}},
 \texttt{\href{mailto:irsic@uw.edu}{irsic@uw.edu}},
 \texttt{\href{mailto:viel@sissa.it}{viel@sissa.it}}

\end{center} 



\vspace{1cm}

\noindent
We investigate constraints on scalar dark matter
 (DM) by analyzing the Lyman-$\alpha$ forest, which probes structure formation
 at medium and small scales, and also by studying its
 cosmological consequences at high and low redshift. For scalar DM that constitutes more than
 30\% of the total DM density, we obtain a lower limit $m \gtrsim
 10^{-21}\, \mathrm{eV}$ for the mass of scalar DM. This implies an upper
 limit on the initial field displacement (or the decay constant for an
 axion-like field) of $\phi \lesssim 10^{16}\, \mathrm{GeV}$. We also
 derive limits on the energy scale of cosmic inflation and establish an
 upper bound on the tensor-to-scalar ratio of $r <  10^{-3}$ in the
 presence of scalar DM. Furthermore, we show that there is very little
 room for ultralight scalar DM to solve the ``small-scale crisis'' of
 cold DM without spoiling the Lyman-$\alpha$ forest results. The constraints
 presented in this paper can be used for testing generic theories that
 contain light scalar fields. 
\vfil

\end{titlepage}

\newpage
\tableofcontents

\section{Introduction}
\label{sec:intro}

The appearance of scalar fields with small masses is ubiquitous in
extensions of the Standard Model of particle physics.
The well-known example is the QCD
axion as a solution to the strong $CP$
problem~\cite{Peccei:1977hh,Weinberg:1977ma,Wilczek:1977pj}.  
String theory even provides a plenitude of ``axion-like fields'' upon 
string compactifications~\cite{Svrcek:2006yi,Douglas:2006es,Arvanitaki:2009fg}.
The smallness of the mass is often attributed to an approximate
shift symmetry, which also suppresses interactions with other fields;
this feature makes the light scalar a good candidate for the dark matter (DM)
of our universe (see~\cite{Ringwald:2012hr} for a review).
Other possible roles of light scalars have also been investigated in the
literature; these include, for example,
explaining the hierarchical flavor structure in the Standard
Model~\cite{Wilczek:1982rv}, 
the electroweak hierarchy~\cite{Graham:2015cka},
and the origin of the baryon asymmetry~\cite{DeSimone:2016bok}.
In any case, however, the weakly interacting light scalars tend to make up
a fraction of the DM, and thus theories containing such fields
can be confronted with cosmological measurements through their scalar DM properties.

The scalar DM behaves similarly to pressureless cold dark matter (CDM)
except for on scales smaller than its de~Broglie wavelength,
where the wave nature of the scalar suppresses structure formation.
In particular for ultralight scalars with $m \sim 10^{-22}\,
\mathrm{eV}$,
the suppression happens on galactic scales and thus drastically modifies
the small-scale structures~\cite{Hu:2000ke} (for a comprehensive review, see~\cite{Hui:2016ltb}).
In this context ultralight scalar DM is also referred to as ``fuzzy
dark matter,'' and has been studied as a possible resolution of the
small-scale discrepancies between N-body simulations of CDM and
observations. 

In this work we investigate the constraints on the mass and fraction of the
scalar DM that are obtained from the Lyman-$\alpha$ forest.  This observable is the main
manifestation of the intergalactic medium, the diffuse matter which fills the space between galaxies, and 
it allows to probe the matter power spectrum at scales and redshifts that are highly complementary to other data sets: the high redshift and small
scales regime.

The Lyman-$\alpha$ forest \cite{Mcquinn} is produced by the absorption of the inhomogeneous distribution of the intergalactic neutral hydrogen along different line of sights to distant quasars~\cite{Viel:2001hd}. It represents, therefore, an extremely useful method for probing the matter power spectrum at small scales, i.e.~$0.5~{\rm Mpc}/h\, \lesssim  \lambda \lesssim 100~{\rm Mpc}/h\, $~\cite{Viel:2013apy,Irsic:2017yje}, since the intergalactic medium displays structure at these scales.
Combining this observable with other observations at larger scales allows also to obtain tight limits on neutrino masses or inflationary models (e.g. \cite{pala,seljak06}).
The scalar DM mass was recently constrained from the Lyman-$\alpha$ forest
in~\cite{Irsic:2017yje,Armengaud:2017nkf} 
for the case where the scalar DM constitutes the entire DM.
In this paper we extend the analysis to cases where the DM consists of
both scalar DM and CDM; this is crucial for constraining general theories
with light scalars which are not necessarily designed to explain the DM
of our universe. 

We then discuss the implications of the Lyman-$\alpha$ bounds for the
nature of the scalar field, and for cosmology.
We evaluate the field range of the scalar, and further derive limits on
the energy scale of cosmic inflation by combining the Lyman-$\alpha$
bounds with constraints from cosmic microwave 
background (CMB) data on isocurvature perturbations.
We also estimate the number of Milky Way satellites arising from scalar DM,
in order to assess the viability of ultralight scalar DM as a solution to
the ``small-scale crisis''.
As a light scalar field generically contributes to DM 
(unless the theory is specifically designed so that the scalar
decays away by the present time, or the scalar is too light such that it
contributes instead to the vacuum energy),
the Lyman-$\alpha$ bounds are shown to provide robust constraints on
theories that contain light scalars.

The paper is organized as follows: in Section \ref{sec:lyman} we review the 
Lyman-$\alpha$ forest constraints of \cite{Irsic:2017yje} and present for the first time
constraints obtained for mixed models; in Section \ref{sec:cosmo}, we quantitatively address
the implications of such a measurement for the early universe and in particular on the initial 
value of the scalar field, isocurvature perturbations, and the tensor-to-scalar ratio. A connection
with the local (late) Universe is provided in Section \ref{astro}, where we compute the number
of Milky Way satellites predicted by the model and compare with the observed number.
Finally, we conclude in Section \ref{sec:conc} with a summary of the main results of this work.
The appendices focus on the exact solution of the Klein--Gordon equation
(Appendix \ref{app:onset}),
on analytic computations of linear density perturbations 
for the mixed model (Appendix \ref{app:lmps}),
and on a comparison between
the constraints presented and those obtained using a simpler yet more approximate method (Appendix \ref{ap:area}).

\section{Lyman-$\alpha$ Forest Constraints}
\label{sec:lyman}

The last decade has seen the emergence of the Lyman-$\alpha$ forest as a
cosmological probe~\cite{Mcquinn}. The light from distant
background sources is scattered on the neutral hydrogen atoms in the
Universe, causing an absorption feature in the observed
spectra, called the Lyman-$\alpha$ forest.  The low-density and
high-redshift intergalactic medium displays a filamentary structure at small and medium scales which is traced by the Lyman-$\alpha$ forest absorption features, and is thus
sensitive to the small scale properties of DM ~\cite{Viel:2013apy,Irsic:2017yje,Armengaud:2017nkf,Irsic:2017ixq,Baur:2017stq}.

We rely on a sample of 100 medium resolution,
high signal-to-noise quasar spectra of the XQ-100 survey \cite{lopez16}, with
emission redshifts $3.5 < z < 4.5$.  A detailed description of
the data and the power spectrum measurements of the XQ-100 survey is 
presented in \cite{Irsic:2017sop}.  Here we repeat the most important
properties of the data and the derived flux power spectrum.  The
spectral resolution of the X-shooter spectrograph is 30-50 km/s,
depending on wavelength.  The flux power spectrum P$_{\rm F}$ (k,z)
has been calculated for a total of 133 $(k,z)$ data points in the
ranges $z=3,3.2,3.4,3.6,3.8,4,4.2$ and 19 bins in $k-$space in the
range 0.003-0.057 s/km (the flux power is indeed estimated in velocity space).  We also use the measurements of the flux
power spectrum of \cite{Viel:2013apy}, at redshift bins
$z=4.2,4.6,5.0,5.4$ and in 10 $k-$bins in the range 0.001-0.08 s/km.
In this second sample the spectral resolution of the quasar absorption
spectra obtained with the MIKE and HIRES spectrographs are about 13.6
and 6.7 km/s, respectively.  As in the analysis of \cite{Viel:2013apy},
a conservative cut is imposed on the flux power spectrum obtained from
the MIKE and HIRES data, and only the measurements with $k >
0.005$   s/km are used to avoid possible systematic uncertainties on
large scales due to continuum fitting (i.e. the removal of long wavelength fluctuations
that are intrinsic of the distant sources and contaminate the large scale power estimates
of $P_{\rm F}(k,z)$, for all the samples considered here these continuum fitting errors
have been estimated).

Compared to XQ-100, the HIRES/MIKE sample has the advantage of probing
smaller scales and higher redshift, where the primordial power spectrum is more linear
and thereby more constraining for the models considered here.  There is a small redshift overlap
between the two samples at $z=4.2$. Since the thermal broadening
(measured in km/s) of Lyman-$\alpha$ forest lines is approximately constant
with redshift, the presence of a cutoff in the matter power spectrum
due to the wave nature of the scalar DM
becomes more prominent in velocity space at high
redshift due to the $H(z)/(1+z)$ scaling between the fixed comoving
length scale set by the scalar DM properties (cf.~(\ref{kJ_eq}),
(\ref{k0.5analytic})) and the corresponding 
velocity scale.  Furthermore, the 1D power spectrum is more sensitive to
the presence of a cutoff compared to the 3D power spectrum, since the value of 1D flux power 
at a given $k_{\rm 1D}$ takes contributions from all the smaller scales $k> k_{\rm 1D}$.

\begin{figure}[t]
  \begin{center}
  \begin{center}
  \includegraphics[width=0.7\linewidth]{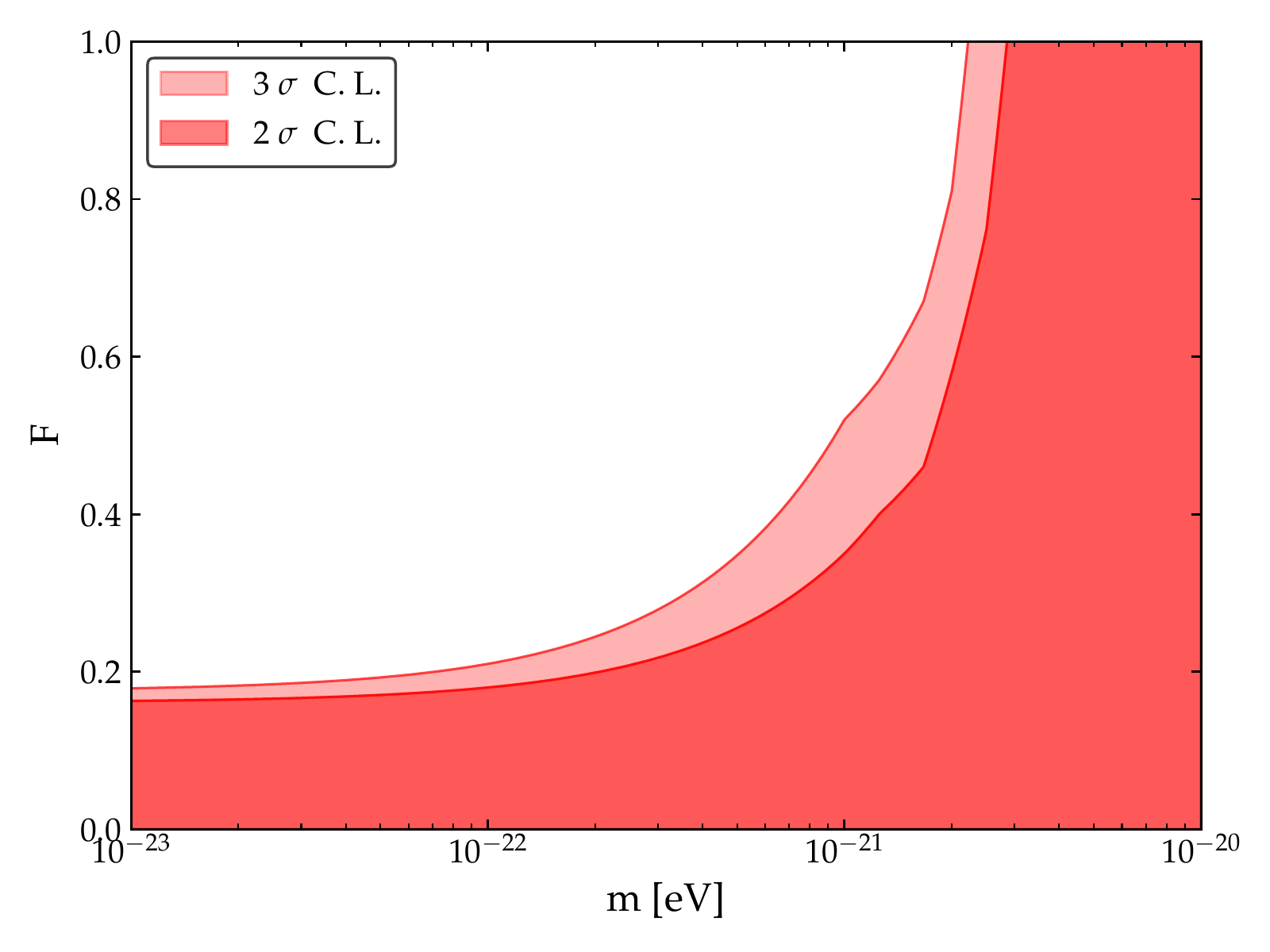}
  \end{center}
   \caption{\emph{Constraints on the scalar DM mass $m$ and
   fraction $F$ of the total DM density in scalar DM obtained from Lyman-$\alpha$ forest data; the two different areas indicate 2 and 3 $\sigma$ confidence levels.
   These results have been obtained for the reference combination of
  data sets described in \cite{Irsic:2017yje}, with a physically
  motivated weak prior on the thermal evolution of the intergalactic
  medium. The regime of $m < 10^{-22}\, \mathrm{eV}$ has been
  extrapolated.}}
  \label{fig:m-f}
  \end{center}
\end{figure}

In terms of simulations, similarly to \cite{Irsic:2017yje}, we model the flux power
spectrum using a set of hydrodynamical simulations performed with the
{\sc{GADGET-III}} code, a modified version of a publicly available {\sc{GADGET-II}}
code \cite{springel05}. The goal of the suite of simulations is to provide a reliable 
template of mock flux power spectra that has to be compared to observations.
Since the flux power spectrum is affected both by astrophysical and cosmological parameters
it is important to properly take them into account and accurately quantify their impact in the likelihood.

We simulate 9 different scalar DM models, with light scalar masses~$m$ of $1$,
$4$ and $15.7 \times 10^{-22}\;\mathrm{eV}$, and
density ratios~$F$ between the scalar DM and total DM (i.e., scalar DM
plus CDM) of $1$, $0.75$ and $0.25$. We also simulate the corresponding $\Lambda$CDM
model, and since the interpolation is done in the
$\alpha=10^{-22}\;\mathrm{eV}/m$, the entire range from $\alpha=0$
($\Lambda$CDM case) to
$\alpha=1$ ($m = 10^{-22}\, \mathrm{eV}$) is covered by interpolation alone. For larger values of
$\alpha$, or equivalently $m < 10^{-22}\, \mathrm{eV}$ a linear
extrapolation is used. In the plane of the DM ratio and the light scalar mass, we
assume that the $\Lambda$CDM model is exact on the axis of $F=0$ and
any $m$. These models were also simulated using the axionCAMB
code \cite{Hlozek:2014lca} to obtain the linear transfer function in the
initial conditions.
At the nonlinear level we do not incorporate the effect of the quantum
pressure, however this should not impact our results for scalar masses down to
$m \sim 10^{-22}\, \mathrm{eV}$, as discussed in
e.g.~\cite{Irsic:2017yje,Schive:2015kza}. 
For even smaller masses the scalar DM fraction becomes small, as we will
shortly see, hence the quantum pressure is also expected to be
negligible there.
If the quantum pressure at the nonlinear level is actually
non-negligible, then it should lead to further suppression of structure
formation; hence the bounds we present for the scalar DM parameters can
be considered as conservative. 

Following \cite{Irsic:2017yje} we vary only $\sigma_8$ (the
normalization of the matter power spectrum) and the slope of the matter
power $n_{\rm eff}$, at the scale of Lyman-$\alpha$ forest (0.005 s/km). Five
different values are considered in the hydrodynamical simulations for
both $\sigma_8$ (in the range of $[0.754, 0.904]$) and $n_{\rm eff}$
(in the range of $[-2.3474,-2.2674]$). These parameters just described
are our cosmological parameters.
There have been several studies in the past
(e.g. \cite{seljak06,McDonald:2004eu,Arinyo-i-Prats:2015vqa}),
that have shown that the Lyman-$\alpha$ forest is really measuring the
amplitude of the linear matter power spectrum, the slope of the power
spectrum, and possibly the effective running, all evaluated at a pivot
scale of around 1-10~${\rm Mpc}/h$. Thus $\sigma_8$ and $n_{\rm eff}$
used are good tracers of what is actually measured.
Given that all our modelling in simulations kept $\Omega_m h^2$ fixed,
$\sigma_8$ can be directly translated into the amplitude of linear 
matter power at the pivot scale (similarly to how $n_{\rm eff}$ was
used). As pointed by~\cite{seljak06}, these matter power amplitude
parameters are equivalent.
The linear matter power only weakly depends on $\Omega_m h^2$, and
moreover, the effects of $\Omega_m$ and $H_0$ on the linear matter power
are already captured in the tracers of the amplitude ($\sigma_8$) and
slope ($n_{\rm eff}$). Therefore the constraints are not sensitive to
the value of $\Omega_m$ nor $H_0$.

Regarding the astrophysical ones, we vary thermal history parameters in the form of the
amplitude ($T_0$) and the slope ($\gamma$) of the intergalactic medium temperature density
relation, usually parameterized as $T=T_0(1+\delta_\mathrm{IGM})^{\gamma-1}$, with $\delta_{\mathrm{IGM}}$ the intergalactic medium overdensity (we refer to \cite{hui97}
for the physical motivation of why the intergalactic medium is
expected to follow the relation above). In the
Monte Carlo Markov Chain (MCMC) runs presented in
Fig.~\ref{fig:m-f} the thermal parameters ($T_0,\gamma$) were assumed to
follow a power-law redshift evolution (e.g. $T_0(z) =
T^A(1+z)^{T^S}$), with weak priors ($T^A \in [0,20000]\;\mathrm{K}$
and $T^S \in [-5,5]$) imposed on the slope and amplitude
of those power-law relations (see reference case
in \cite{Irsic:2017yje}). Thermal evolution described by such
power-laws agrees well with the temperature measurements found in the
literature (\cite{Becker2011,boera2014}). However, even if more conservative temperature
evolution with redshift was allowed in the MCMC runs, the MCMC
constraints are expected to become weaker by only an order of unity
(as was the case for $F =1$ in \cite{Irsic:2017yje}; 
see $T_0(z)$ bins case in Table I).
The conservative approach allowed $T_0(z)$
to vary independently in each redshift bin, but prevented un-physical
jumps in temperature (jumps with $>5000\;\mathrm{K}$ were not allowed
between consecutive redshift bins).

We also vary the timing of the instantaneous reionization
model $z_{\rm rei}$. As in \cite{Irsic:2017yje}, three values for
each of these parameters are considered, in the regime based on recent
observational results. The thermal history is supposed to be the most important contaminant
since a hotter medium in general tends to produce a smoother flux
distribution and a flux power with less substructure at small scales,
like warm DM or ultralight scalar DM models
do. However, the redshift evolution of thermal and cosmological effects is very different and the wide range
explored by our data allows to break the degeneracies between the parameters in a very effective way.
We further consider ultraviolet (UV) fluctuations of the ionizing background, which could be particularly important at high redshift and 
build a refined power spectrum template that  incorporates this effect. The amplitude of this effect is let free and is described by the parameter
f$_{\rm UV}=[0,1]$, which we marginalize over in the final constraints.
We do not consider here temperature fluctuations which have been advocated as potentially mimicking the 
presence of a cutoff at small scales in \cite{Hui:2016ltb}. However, according to sophisticated and recent hydrodynamical simulations
these effects appear to happen at large and not small scales \cite{cen09,aloisio15}.
A comprehensive treatment of spatial UV and temperature fluctuations would require computationally prohibitive
radiative transfer calculations in large volumes and it is beyond the analysis performed here.

With the models of the flux power spectra obtained from the hydrodynamical
simulations, we establish a sparse grid of points in the parameter
space and by using linear interpolation between the grid points we obtained
predictions for the quantity $P_{\rm F}(k,z,{\bf{p}})$, with ${\bf{p}}$ a vector containing all the parameters described in the analysis, by performing a Taylor expansion for the desired models in a much finer grid for the highly multidimensional parameter space.   We refer to \cite{v06} for a more detailed description of the basic idea of this approach.
We use an MCMC code in order to estimate the parameter constraints, and the cast results in terms
of mass and fraction of scalar DM have been obtained by marginalizing over the whole set of other parameters. The results of the MCMC  for the mixed models are shown in Figure~\ref{fig:m-f}, using the  reference analysis of  \cite{Irsic:2017yje} which relies on all the data sets and the assumption that the thermal state evolution
follows a power-law, without any prior on the cosmological parameters.

When the scalar DM constitutes the entire DM, i.e. $F = 1$, the
Lyman-$\alpha$ forest data yields a lower bound on the scalar mass of $m
\gtrsim 10^{-21}\, \mathrm{eV}$, as was also shown
in~\cite{Irsic:2017yje}. On the other hand, the Lyman-$\alpha$ forest
becomes insensitive to scalar DM at $F \lesssim 0.2$,
which reflects the fact that the matter power spectrum is only mildly
suppressed by scalar DM with such small fraction, no matter how light it is.
(This is explicitly shown in Appendix~\ref{app:lmps}
through analytic computations of the linear matter power spectrum.)
Although the regime of $m < 10^{-22}\, \mathrm{eV}$ is only explored
through extrapolation of our simulation models, we do not expect the
bound in this regime to change significantly even with actual simulations,
as the finite size of the error bars on the measured flux
power spectrum makes it difficult to detect the mild suppression
of the matter power.

Constraints on small-scale properties of the DM for ultra-light bosons, resonantly produced sterile neutrinos and thermal relics,
also in mixed cold and warm models have also recently been presented in \cite{Armengaud:2017nkf,Irsic:2017ixq,Baur:2017stq}, with results
that are in overall agreement and exclude models with a relatively strong suppression of power to alleviate the small-scale problems of CDM.

The results of this paper are based on the full MCMC analysis outlined
in this Section. However, it is possible to get a simpler, although less accurate, grasp of the Lyman-$\alpha$ forest constraints by applying an intuitive method dubbed as \emph{area criterion}~\cite{Murgia:2017lwo}. This method 
and its comparison with the MCMC results are discussed in Appendix~\ref{ap:area}.

\section{Cosmological Implications}
\label{sec:cosmo}

We now move on to discuss the implications of the Lyman-$\alpha$
constraints for the scalar field, and for the early universe. 
After evaluating the field range of the scalar, we 
compute the isocurvature perturbations in the scalar DM
density sourced during cosmic inflation.
Combining with CMB constraints on DM isocurvature,
we also derive bounds on the inflation scale.

\subsection{Initial Displacement from the Vacuum}
\label{subsec:abundance}

A light scalar field stays frozen at its initial field value in the
early universe. Hence any initial displacement from the potential
minimum gives rise to a scalar DM density in the later universe.
We consider such an initial vacuum misalignment to be the main source of
the density, 
and also suppose the scalar mass to be time-independent 
(unlike the QCD axion whose mass depends on the cosmic temperature).
Then the scalar can collectively be described by a homogeneous
Klein--Gordon equation in a FRW background universe,
\begin{equation}
 \ddot{\phi} + 3 H \dot{\phi} + m^2 \phi = 0,
\label{homoKG}
\end{equation}
where an overdot denotes a derivative in terms of the cosmological time,
and $H = \dot{a} / a$.
The homogeneous scalar field forms a perfect fluid with an energy
density and pressure of 
\begin{equation}
 \rho_\phi = \frac{1}{2} \left(\dot{\phi}^2 + m^2 \phi^2 \right),
  \quad
 p_\phi = \frac{1}{2} \left(\dot{\phi}^2 - m^2 \phi^2 \right).
\label{rhoandp}
\end{equation}
We denote the initial displacement of the scalar field from its
potential minimum by~$\phi_\star$. 
In the early universe when $H \gg m$, the scalar field is frozen
at~$\phi_{\star}$ due to the Hubble friction, and thus 
contributes to the vacuum energy.
On the other hand in the later universe when $H \ll m$,
the scalar undergoes harmonic oscillations along the quadratic potential
and behaves as pressureless matter.
Thus the scalar densities in the two epochs are written as
\begin{numcases}{\rho_\phi = }
    \frac{1}{2} m^2 \phi_{\star}^2
       & when $H \gg m$, \label{rho_early} \\
    \frac{1}{2} m^2 \phi_{\star}^2 \left(\frac{a_{\mathrm{osc}}}{a}\right)^3
       & when $H \ll m$. \label{rho_late}
\end{numcases}
These asymptotic behaviors smoothly connect to each other at
around $H \sim m$.
Here $a_{\mathrm{osc}}$ represents the scale factor at the `onset' of
the scalar oscillation;
the explicit value of~$a_{\mathrm{osc}}$ is chosen such that 
the scalar density in the asymptotic future
matches with the expression~(\ref{rho_late}).
We also denote quantities measured at $a = a_{\mathrm{osc}}$ by the
subscript~``osc''.

We are interested in ultralight scalars that start oscillating in the
radiation-dominated epoch, instead of during times prior to reheating. 
The exact solution of the Klein--Gordon equation in a radiation-dominated
background is given in Appendix~\ref{app:onset}, 
where the ratio between the mass and Hubble rate at $a =
a_{\mathrm{osc}}$ is shown to take the value of
\begin{equation}
 \frac{m^2}{H_{\mathrm{osc}}^2} = 
  \left( \frac{8}{\pi }  \right)^{4/3}
  \left[ \Gamma \left(\frac{5}{4} \right) \right]^{8/3}
  \approx 2.68.
\label{2.68}
\end{equation}
Since the total energy density, and hence the Hubble rate, of a radiation-dominated
universe are related to the cosmic temperature by
\begin{equation}
 \rho_{\mathrm{r}} =  3 M_p^2 H^2  = \frac{\pi^2}{30}g_* T^4,
\label{rho_RD}
\end{equation}
with $M_p = (8 \pi G)^{-1/2} $
being the reduced Planck mass,
the temperature at $a = a_{\mathrm{osc}}$ is obtained as
\begin{equation}
 T_{\mathrm{osc}} \approx 0.5 \, \mathrm{keV}
  \left(\frac{g_{*\mathrm{osc}}}{3.36}\right)^{-1/4}
  \left(\frac{m}{10^{-22}\, \mathrm{eV}}  \right)^{1/2}.
  \label{Tosc}
\end{equation}
Thus for instance, a scalar with $m = 10^{-22}\, \mathrm{eV}$
starts oscillating when the cosmic temperature drops to
$T_{\mathrm{osc}} \approx 0.5 \, \mathrm{keV}$.
Moreover, the scalar would be oscillating at
matter-domination equality as long as $m \gg 10^{-28}\, \mathrm{eV}$.

Using that the entropy of the universe is conserved since the radiation-dominated
epoch, the scalar density today can be expressed in terms of the entropy density~$s$ as
\begin{equation}
 \rho_{\phi 0 } = \frac{1}{2} m^2 \phi_{\star}^2
  \, \frac{s_0}{s_{\mathrm{osc}}},
\label{rhophi0}
\end{equation}
where the subscript~``$0$'' denotes quantities in the present universe.
Here, $s_{\mathrm{osc}}$ is written in terms of $H_{\mathrm{osc}}$
using~(\ref{rho_RD}) as
\begin{equation}
 s_{\mathrm{osc}}
= \frac{2 \pi^2}{45}g_{s* \mathrm{osc}} T_{\mathrm{osc}}^3
= \frac{2 \pi^2}{45}
  g_{s*\mathrm{osc}}
  \left(\frac{90}{\pi^2}
 \frac{ M_p^2 H_{\mathrm{osc}}^2}{g_{*\mathrm{osc}}}
  \right)^{3/4}  .
\label{sosc}
\end{equation}

\begin{figure}[t]
  \begin{center}
  \begin{center}
  \includegraphics[width=0.7\linewidth]{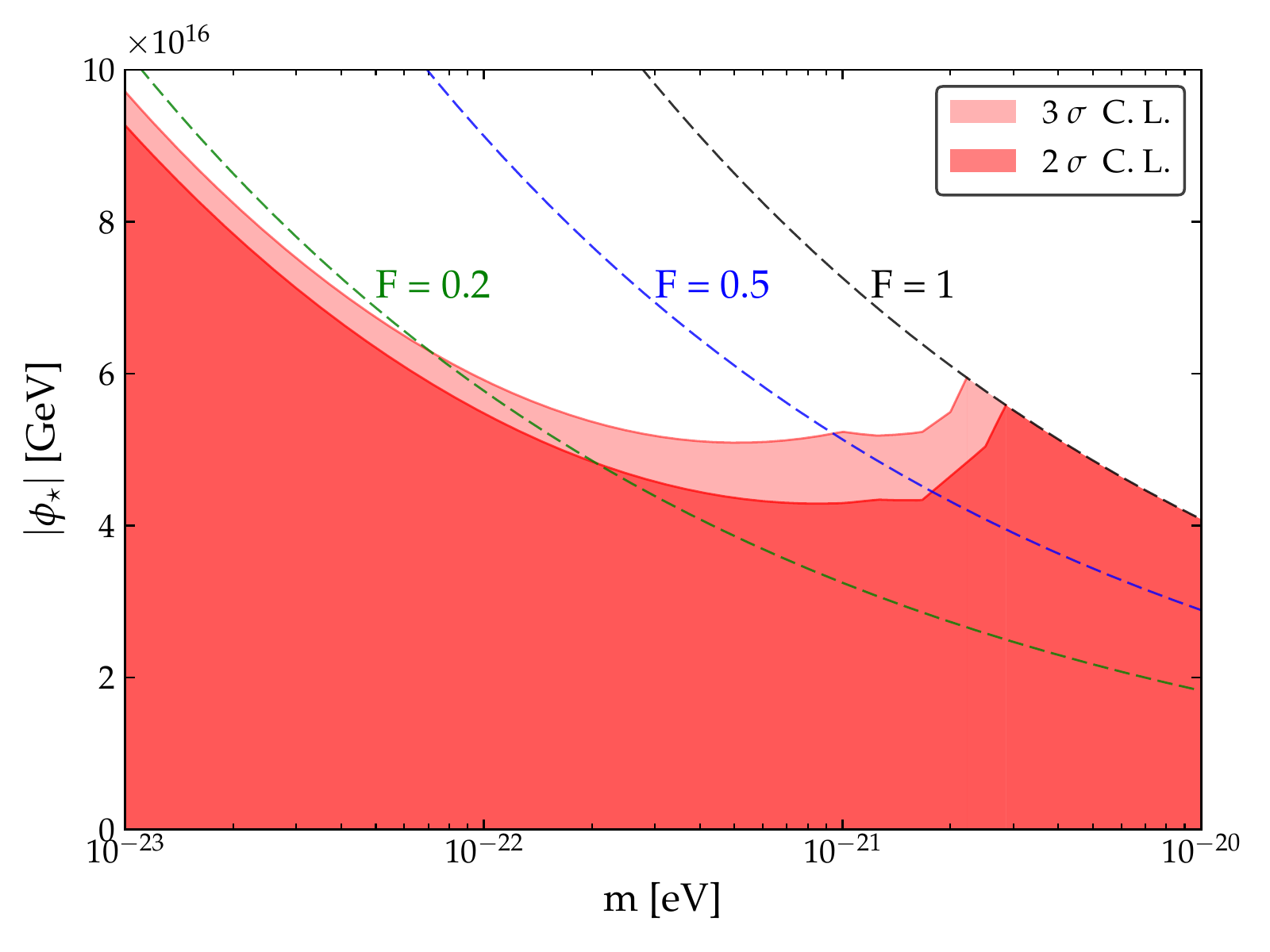}
  \end{center}
   \caption{\emph{Upper limits on the initial displacement of the scalar
   field~$\phi_\star$ from Lyman-$\alpha$ forest data (2 and 3~$\sigma$
   C.L.) as a function of the scalar DM mass $m$.
   The colored dashed lines show contours of constant scalar DM fraction~$F$.}}
  \label{fig:phi-f}
  \end{center}
\end{figure}

Thus the present-day scalar DM abundance is obtained by combining (\ref{2.68}),
(\ref{rhophi0}), and (\ref{sosc}).
Expressing it in terms of the ratio to the CDM density measured by {\it
Planck}~\cite{Ade:2015xua}, 
$\Omega_{\mathrm{c}} h^2 = 0.1186 \pm 0.0020$
(68\% C.L., TT+lowP+lensing), one finds\footnote{Note that $F$ is
defined in this paper as the density ratio between scalar DM and total
DM (i.e., scalar DM plus CDM). Here we are identifying the measured CDM
density with the total DM density.}
\begin{equation}
 F \equiv \frac{\Omega_{\phi}}{\Omega_{\mathrm{c}}} \approx 0.6 \, 
  \left(\frac{g_{*\mathrm{osc}}}{3.36}\right)^{3/4}
  \left(\frac{g_{s*\mathrm{osc}}}{3.91}\right)^{-1}
  \left( \frac{\phi_{\star}}{10^{17} \, \mathrm{GeV}} \right)^2
   \left( \frac{m}{10^{-22} \, \mathrm{eV}} \right)^{1/2}.
\label{Omega_phi}
\end{equation}
Using this relation, the Lyman-$\alpha$ constraints on ($m$, $F$),
cf. Figure~\ref{fig:m-f}, are
translated into bounds on the scalar parameters ($m$, $\phi_\star$).
In Figure~\ref{fig:phi-f} we plot the 2 and 3$\sigma$ limits on ($m$,
$\phi_\star$) from the Lyman-$\alpha$ forest data, where the shaded regions
indicate the allowed parameter space. 
Here we have set $g_{*\mathrm{osc}} = 3.36$, $g_{s*\mathrm{osc}} = 3.91$, since
$T_{\mathrm{osc}} \ll 1 \, \mathrm{MeV}$ for the displayed masses
(cf.~(\ref{Tosc})). 
The dashed lines in  Figure~\ref{fig:phi-f} indicate the contours of constant fraction $F$.
We find that the initial displacement is bounded from above as
$\abs{\phi_\star} \lesssim 10^{16}\, \mathrm{GeV}$ for most values of
the mass.
For masses of $m \gtrsim 10^{-21}\, \mathrm{eV}$,
$\phi_\star$ is constrained mainly by the requirement
that the scalar should not lead to overabundance of DM (i.e. $F
\leq 1$).
On the other hand for $m \lesssim
10^{-21}\, \mathrm{eV}$, the Lyman-$\alpha$ forest gives the strongest
constraint.
As one goes to even smaller masses $m \lesssim 10^{-22}\,
\mathrm{eV}$, the Lyman-$\alpha$ bound on~$\phi_\star$ weakens since the
scalar DM density~$F$ decreases. There the Lyman-$\alpha$ bound closely
follows the $F = 0.2$ contour, allowing $\phi_\star$ to take larger values.

\subsection{Isocurvature Perturbation and Inflation Scale}
\label{subsec:iso}

The ultralight scalar acquires super-horizon field fluctuations during
cosmic inflation.\footnote{This is not necessarily the case if
the scalar arises {\it after} inflation as a 
pseudo Nambu--Goldstone boson of a spontaneously broken global U(1) symmetry.
However in such cases, topological defects are produced, which would
overclose the universe (unless the number of degenerate vacua along the
bottom of the Mexican hat potential is
$N=1$)~\cite{Vilenkin:1982ks,Sikivie:1982qv,Linde:1990yj}.
For a recent discussion, see also~\cite{Visinelli:2017imh}.
Thus in this Section we suppose the scalar field to have
existed already during inflation.}
As a consequence, the initial field value~$\phi_\star$
possesses fluctuations with a power spectrum of 
\begin{equation}
 \mathcal{P_{\delta \phi_\star}} (k) = 
  \left( \frac{H_k}{2 \pi} \right)^2,
\label{phi-fluc}
\end{equation}
where $H_k$ represents the Hubble rate during inflation when the comoving
wave number~$k$ exits the horizon.
Since the scalar does not dominate the universe until the
matter-radiation equality, its field fluctuations lead to
isocurvature perturbations.

From~(\ref{rhophi0}) the scalar DM density depends on the
initial field value as $\rho_\phi \propto \phi_\star^2$, therefore
the density fluctuates as
\begin{equation}
 \frac{\delta \rho_\phi }{\rho_\phi} = 2 \frac{ \delta \phi_\star }{ \phi_\star},
\end{equation}
up to linear order in the field fluctuations.
Identifying this with the scalar DM isocurvature perturbation~$S_{\phi \gamma}$
using (\ref{phi-fluc}),
and further multiplying with the DM fraction yields the effective CDM
isocurvature power spectrum,
\begin{equation}
 \mathcal{P}_{\mathrm{c} \gamma} (k) =
  F^2
  \mathcal{P}_{\phi \gamma} (k) =
    \left( \frac{F  H_k}{\pi
     \phi_\star} \right)^2 .
\label{effCiso}
\end{equation}
Given that the Hubble rate during inflation is nearly constant,
the isocurvature spectrum is nearly scale-invariant.
Moreover, the scalar~$\phi$ does not contribute to curvature
perturbations and hence there is no correlation between the 
isocurvature and curvature perturbations.

Since the scalar DM compatible with the Lyman-$\alpha$ analysis 
behaves similarly to CDM on large scales,
the CMB constraints on CDM isocurvature perturbations also apply to scalar DM.
Parameterizing the isocurvature power spectrum in terms of the curvature
power as
\begin{equation}
 \mathcal{P}_{\mathrm{c} \gamma} (k)
  = \frac{\beta_{\mathrm{iso}} (k)}{1 - \beta_{\mathrm{iso}} (k)}
  \mathcal{P}_\zeta (k),
\end{equation}
uncorrelated and scale-invariant CDM isocurvature
is constrained by {\it Planck}~\cite{Ade:2015lrj} 
at the pivot scale $k_{\mathrm{piv}} / a_0 = 0.05 \, \mathrm{Mpc}^{\text{-}1} $ as
\begin{equation}
 \beta_{\mathrm{iso}} (k_{\mathrm{piv}}) < 0.038 \quad
  \text{(95\% C.L., TT, TE, EE+lowP)},  
\end{equation}
with $\mathcal{P}_\zeta (k_{\mathrm{piv}}) \approx 2.2 \times
10^{-9}$.

\begin{figure}[t]
  \begin{center}
  \begin{tabular}{lr}
  \includegraphics[width=0.7\linewidth]{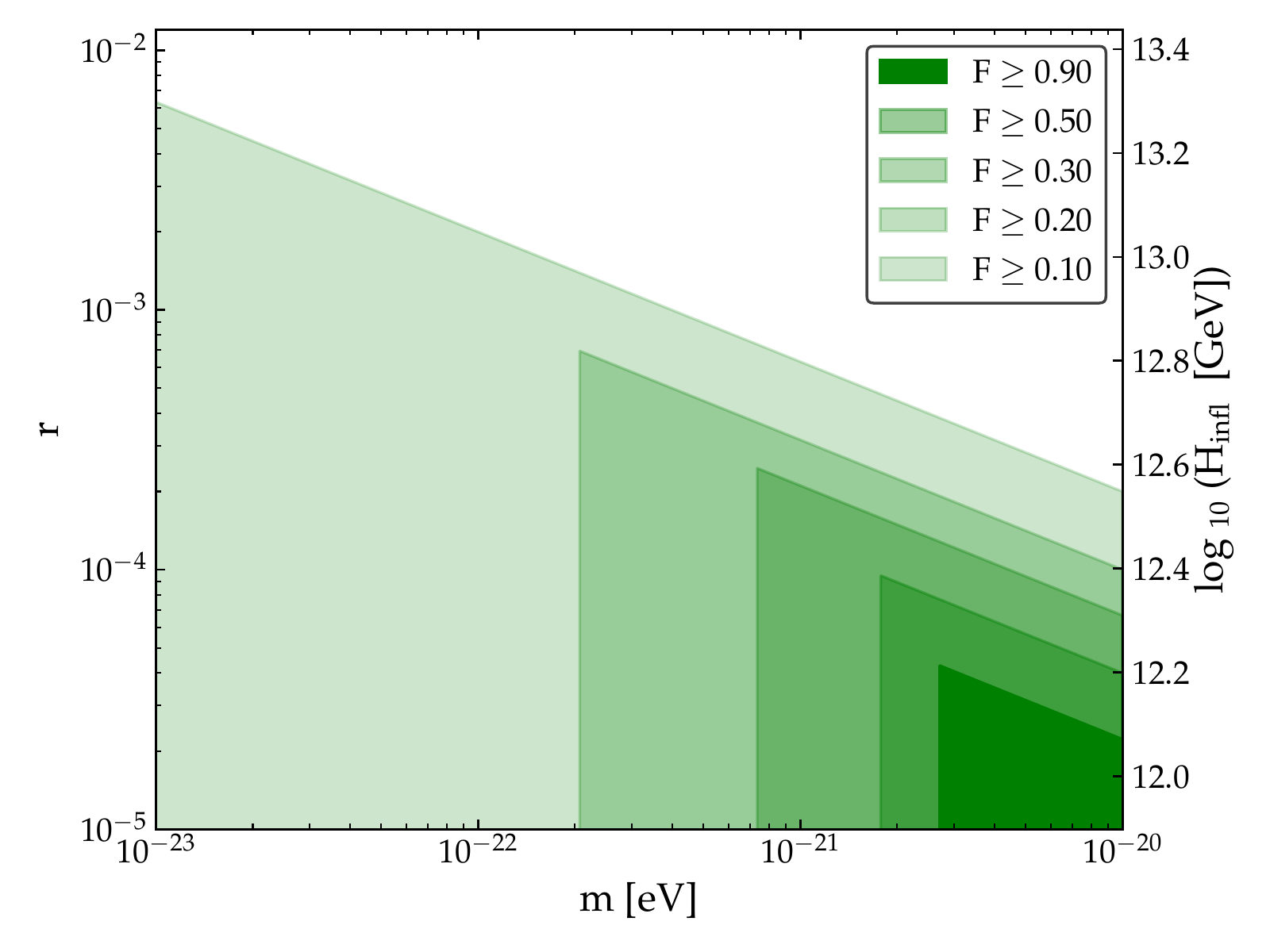}
  \end{tabular}
   \caption{\emph{Upper bound on the inflation scale~$H_{\mathrm{inf}}$ and
   tensor-to-scalar ratio~$r$ at the pivot scale $k_{\mathrm{piv}} / a_0
   = 0.05 \, \mathrm{Mpc}^{\text{-}1} $, 
   as a function of the scalar DM mass~$m$ (2$\sigma$ C.L.). 
   Differently colored regions represent the allowed parameter space
   when the scalar DM constitutes a certain fraction~$F$ of the total DM.}}
  \label{fig:m-r}
  \end{center}
\end{figure}

The {\it Planck} upper bound on the isocurvature translates
into a bound on the inflation scale through~(\ref{effCiso});
eliminating~$\phi_\star$ using~(\ref{Omega_phi}),
we obtain an upper limit on the Hubble rate when the pivot scale leaves
the horizon as
\begin{equation}
 H_{k_{\mathrm{piv}}} <
  4 \times 10^{12}\, \mathrm{GeV} 
  \left(\frac{g_{*\mathrm{osc}}}{3.36}\right)^{-3/8}
  \left(\frac{g_{s*\mathrm{osc}}}{3.91}\right)^{1/2}
  F^{-1/2}
   \left( \frac{m}{10^{-22} \, \mathrm{eV}} \right)^{-1/4}.
\label{Hinfbound}
\end{equation}
This can also be expressed as a bound on the tensor-to-scalar
ratio,\footnote{Here we assume that the sound speed of the tensor
fluctuations is unity.}
\begin{equation}
 r(k) = \frac{\mathcal{P}_T (k)}{\mathcal{P}_\zeta (k)}
  = \frac{1}{\mathcal{P}_{\zeta } (k)}\frac{2 H_k^2}{\pi^2 M_p^2},
\label{ts-ratio}
\end{equation}
as
\begin{equation}
 r (k_{\mathrm{piv}}) <
  2 \times 10^{-4} \, 
  \left(\frac{g_{*\mathrm{osc}}}{3.36}\right)^{-3/4}
  \left(\frac{g_{s*\mathrm{osc}}}{3.91}\right)
  F^{-1}
   \left( \frac{m}{10^{-22} \, \mathrm{eV}} \right)^{-1/2}.
\label{rbound}
\end{equation}
Alternatively, in terms of $m$ and $\phi_\star$, the bound is written as
\begin{equation}
 r (k_{\mathrm{piv}}) <
  4 \times 10^{-4} \, 
  \left(\frac{g_{*\mathrm{osc}}}{3.36}\right)^{-3/2}
  \left(\frac{g_{s*\mathrm{osc}}}{3.91}\right)^2
  \left( \frac{m}{10^{-22} \, \mathrm{eV}} \right)^{-1}
  \left( \frac{\phi_\star}{10^{17} \, \mathrm{GeV}} \right)^{-2}.
\label{rbound2}
\end{equation}
These constraints become weaker for a smaller~$m$.
On the other hand, the Lyman-$\alpha$ forest sets a lower bound on~$m$.
Thus by combining the Lyman-$\alpha$ and CMB constraints, an upper bound
on the inflation scale can be obtained.
This is presented in Figure~\ref{fig:m-r}, where
each colored region represents the values allowed for the scalar DM
mass~$m$ and the tensor-to-scalar ratio~$r$, or the inflation
scale~$H_{\mathrm{inf}}$,
when the scalar DM constitutes a certain fraction~$F$ of the total DM.
Here we combined the $2\sigma$~limit on scalar DM from the
Lyman-$\alpha$ forest analysis (cf.~Figure~\ref{fig:m-f}) with the
{\it Planck} $2\sigma$~limit on isocurvature perturbations
(i.e.~(\ref{rbound}) with 
$g_{*\mathrm{osc}} = 3.36$, $g_{s*\mathrm{osc}} = 3.91$).
The former sets the left boundaries of each region, and the latter the
upper boundaries.\footnote{Isocurvature perturbations can
also impact the Lyman-$\alpha$ 
forest~\cite{Beltran:2005gr}, thus for a rigorous treatment,
the isocurvature should also be included in the Lyman-$\alpha$ analyses. 
However since the scalar DM isocurvature is nearly scale-invariant, 
its effect on the Lyman-$\alpha$ should be tiny;
hence here we simply combine the result of Section~\ref{sec:lyman}
with the ${\it Planck}$ limit.}
One clearly sees that scalar DM is incompatible with an observably
large~$r$,
with the upper limits on $r$ becoming stronger for a larger~$F$.
In particular if the scalar DM constitutes more than 20\% of the total
DM, the tensor-to-scalar ratio would be as low as $r < 10^{-3}$.
This in turn suggests that any detection of primordial gravitational
waves in the near future would rule out scalar DM produced from a vacuum
misalignment as the main component of DM.\footnote{However we should also
remark that there has been attempts to make light scalar DM consistent
with high-scale inflation by adding further ingredients.
One such example is an axion-like field with a time-dependent decay
constant~\cite{Linde:1990yj,Linde:1991km,Higaki:2014ooa,Chun:2014xva,Fairbairn:2014zta,Kobayashi:2016qld}.} 
We will also illustrate this point in Figure~\ref{fig:summary}, where 
contours of the upper bounds of~$r$~(\ref{rbound2}) are shown 
on the ($m$, $\phi_\star$) plane;
regions above the contour would be excluded if $r$ is detected at the
displayed value.

\subsection{Comments on Axion-like Fields}
\label{subsec:axion}

Scalar fields with approximate continuous shift symmetries, 
often referred to as axion-like fields, have been studied as an
ultralight DM candidate~\cite{Arvanitaki:2009fg,Ringwald:2012hr,Hui:2016ltb}.
These models are typically described by an action with a periodic
potential of
\begin{equation}
 S = \int d^4 x \sqrt{-g}
  \left[
   -\frac{1}{2} f_a^2  g^{\mu \nu} \partial_\mu \theta \partial_\nu \theta
 - m^2 f_a^2 \left( 1 - \cos \theta  \right)
  \right],
\end{equation}
with $f_a$ being an ``axion decay constant''.
When focusing on the vicinity of a minimum $\theta = 0$,
the potential is expanded as 
\begin{equation}
 V(\theta) = m^2 f_a^2 \left( 1 - \cos \theta  \right) =
  \frac{1}{2} m^2 f_a^2 \theta^2 + \mathcal{O} (\theta)^4.
\label{cosine-pot}
\end{equation}
Thus our analyses in the previous Sections apply to
axion-like fields by replacing~$\phi$ with a product of the decay
constant and the angle, i.e.,
\begin{equation}
 \phi \to f_a \, \theta.
\end{equation}
In particular, if the initial misalignment angle $\theta_\star$ is of
order unity, then the bounds on~$\phi_\star$ directly translates into
bounds on the decay constant~$f_a$.

However we should also remark that the discussions are modified if the
initial angle is as large as $\abs{\theta_\star } > 1$;
then the expansion~(\ref{cosine-pot}) breaks down and the axion
potential can no longer be approximated by a quadratic.
The non-quadratic nature of the axion potential would lead to
(i) an increase in the final axion DM density due
to a delayed onset of the scalar
oscillation~\cite{Turner:1985si,Bae:2008ue}, and (ii) a significant 
enhancement of the axion isocurvature due to a nonuniform onset of the oscillation,
giving much stronger bounds on the inflation
scale~\cite{Lyth:1991ub,Strobl:1994wk,Kobayashi:2013nva}.
We also note that, besides such anharmonic initial conditions,
a time-dependent $m$ (e.g. of the QCD axion)
or $f_a$ (e.g.~\cite{Linde:1990yj,Linde:1991km,Higaki:2014ooa,Chun:2014xva,Fairbairn:2014zta,Kobayashi:2016qld}) can also modify the cosmological
evolution of the axion.

\section{Implications for the Number of Milky Way Satellites}
\label{astro}

In this Section we discuss the astrophysical implications of an
ultralight scalar DM scenario: the possibility, in this framework, of alleviating the ``small-scale crisis'' of the standard CDM paradigm. In particular, we focus on the well known {\emph{missing satellite}} problem~\cite{Klypin:1999uc,Moore:1999nt}. 

It is nowadays well established that DM models described by suppressed matter power spectra may be able to relax the tensions present in the standard CDM context at sub-galactic scales, i.e.~the discrepancy between the observed number of dwarf galaxies within the Milky Way (MW) virial radius and the number of MW substructures predicted by cosmological $N$-body simulations, assuming the standard CDM model.
Current analyses claim, for instance, that thermal warm DM candidates with masses between 2 and 3 keV can induce a suppression in the corresponding matter power spectra such that this tension vanishes or is greatly reduced (e.g.,~\cite{lovell,lovell2}).
It is thus interesting to investigate the implications of our scenario
at sub-galactic scales, in order to check if the ($m$, $F$)-combinations which have been found to be in agreement with Lyman-$\alpha$ forest data are also capable of solving/alleviating the {\emph{missing satellite}} problem.

An accurate calculation of the number of substructures $N_{\rm sub}$
with ultralight scalar DM would require high-resolution $N$-body
simulations, which is beyond the scope of this paper.
Here we instead make a rough estimate of~$N_{\rm sub}$ using the
following analytical expression for the number of subhalos,
\begin{equation}\label{eq:subMF}
 \frac{{\rm d}N_{\rm sub}}{{\rm d}M_{\rm sub}} = \frac{1}{44.5}\frac{1}{6\pi^2}\frac{M_{\rm halo}}
 {M^2_{\rm sub}}\frac{P(1/R_{\rm sub})}{R^3_{\rm sub}\sqrt{2\pi(S_{\rm sub}-S_{\rm halo})}},
\end{equation}
which was introduced in~\cite{Schneider:2014rda,Schneider:2016uqi}
based on the conditional mass function normalized to the
$N$-body simulation results.\footnote{Notice that the use of this
procedure is supported by the recent analysis performed in
Ref.~\cite{Murgia:2017lwo}, where the accuracy of the theoretical
predictions (Eq.\eqref{eq:subMF}) has been checked against a large suite
of $N$-body simulations.}  
Here $R_{\rm sub}$, $M_{\rm sub}$ and $S_{\rm sub}$ are radius, mass and variance of a given subhalo, while $M_{\rm halo}$ and $S_{\rm halo}$ are the mass and the variance of the main halo, defined as follows:
\begin{equation}\label{eq:mass_var}
 S_i = \frac{1}{2\pi^2} \int\limits_0^{1/R_i} {\rm d}k\ k^2P(k), ~~~~~ M_i = \frac{4\pi}{3}\Omega_m\rho_c(cR_i)^3, ~~~~~ c=2.5,
\end{equation}
with $P(k)$ being the linear power spectrum of a given model computed at
redshift $z=0$, and $\rho_c$ the critical density today.

For a given MW halo mass, e.g. $M_{\rm halo} = 1.7 \cdot
10^{12}~M_\odot/h$~\cite{Lovell:2013ola}, we can now obtain the number
of subhalos $N_{\rm sub}$ with masses $M_{\rm sub} \geq 10^8~M_\odot/h$
predicted by different parameterizations of
the scalar DM scenario (i.e.~by different combinations of $m$ and
$F$). This is done simply by integrating Eq.~\eqref{eq:subMF}. 
Here, note that a different choice of $M_{\rm halo}$ mainly leads to an
overall shift in $N_{\rm sub}$ for all DM scenarios,
as is seen from the $M_{\rm halo}$-dependence of Eq.~\eqref{eq:subMF}.
Hence instead of studying the absolute value of~$N_{\mathrm{sub}}$, which
is sensitive to the MW halo mass,
we firstly would like to focus on the relative suppression
of~$N_{\mathrm{sub}}$,
i.e., the ratio between $N_{\mathrm{sub}}$ for cases with and without
scalar DM. 
As a benchmark value for the relative suppression of~$N_{\mathrm{sub}}$, 
we take the thermal warm DM with masses in
the interval between 2 and 3 keV as reference models, 
and consider the mixed scalar DM and CDM scenarios
that yield similar relative suppressions
to be able to solve the {\emph{missing satellite}} problem.

\begin{figure}[t]
  \begin{center}
  \begin{center}
  \includegraphics[width=0.7\linewidth]{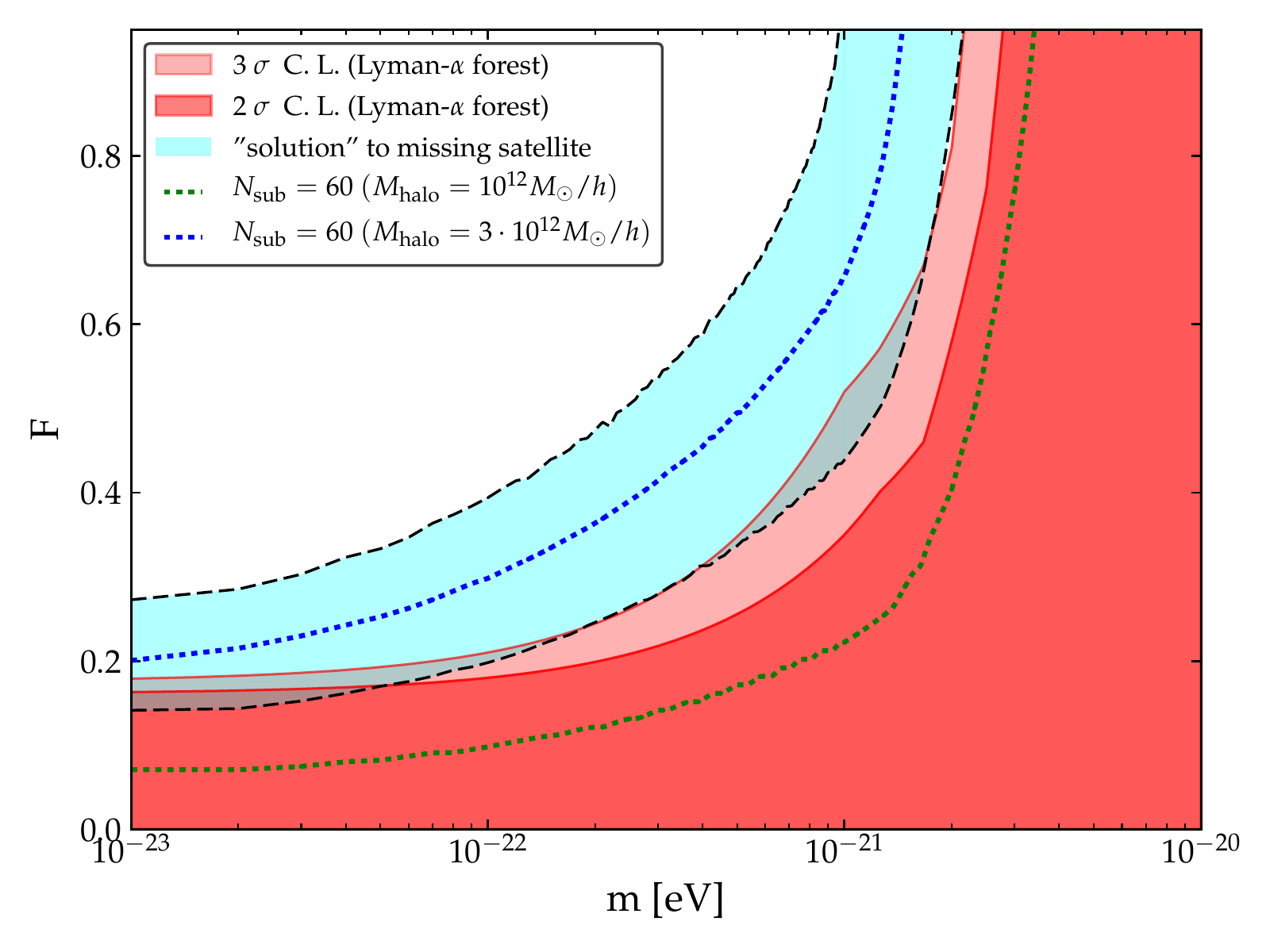}
  \end{center}
   \caption{\emph{The comparison between the constraints on the scalar
   DM parameter space from the Lyman-$\alpha$ forest data analysis at 2
   and 3$\sigma$ (red regions, see Fig.~\ref{fig:m-f}), and the region
   capable of ``solving'' the missing satellite problem (cyan region bounded
   by dashed lines). The green and blue dotted lines refer to models
   which predict $N_{\mathrm{sub}}=60$, 
   when choosing $M_{\rm halo} = 10^{12} M_{\odot} / h$ and $M_{\rm
   halo} = 3 \cdot 10^{12} M_{\odot} / h$, respectively.}}  
  \label{fig:MissingSat}
  \end{center}
\end{figure}

The parameter window for ($m$, $F$) of scalar DM where the
number of subhalos lies within the ``solving'' range is shown in 
Fig.~\ref{fig:MissingSat} as the cyan shaded area bounded by dashed lines.
For reference, when using $M_{\rm halo} = 1.7 \cdot 10^{12}~M_\odot/h$,
the aforementioned computation gives
the number of subhalos with CDM only as $N_{\rm
sub} = 158$, while $20 \leq N_{\rm sub} \leq 60$ for the reference warm DM models.
(However, as we explained, these absolute values are irrelevant when
focusing on the relative suppression of~$N_{\rm sub}$.)
The red shaded areas in Fig.~\ref{fig:MissingSat} represent the 2 and
3$\sigma$ contours from the Lyman-$\alpha$ forest data analysis,
discussed in Section~\ref{sec:lyman}. As one can easily see from the
plot, there is very little room for simultaneously satisfying these
constraints and solving the {\emph{missing satellite}} problem. 

Let us also discuss the effects of the observational
uncertainties in the MW halo mass;
for instance, a recent comprehensive dynamical analysis of redshifts and
distances of 64 dwarf galaxies around the MW has led to $M_{\rm halo} =
2.8 \cdot 10^{12}~M_\odot$~\cite{2017Peebles}.
The detailed value of $M_{\rm halo}$ does matter when focusing
instead on the absolute value of~$N_{\mathrm{sub}}$,
and past studies such as~\cite{Kennedy:2013uta} have pointed out the 
degeneracy between the MW halo mass and the DM
parameters required for relaxing the {\emph{missing satellite}} problem.

In order to take into account these issues, we have iterated the same
analysis with different input values for~$M_{\rm
halo}$,
and compared the corresponding predictions to a fixed satellite number 
$N_{\rm sub} = 60$;
this value is chosen as a sum of the 
11 MW classical satellites and the 15 ultra-faint satellites from SDSS, 
with the latter value multiplied by a numerical factor which accounts
for the limited sky coverage of the
survey~\cite{Murgia:2017lwo,Schneider:2016uqi,Polisensky:2010rw}.
For $M_{\rm halo} = 1.7 \cdot 10^{12}~M_\odot/h$,
the number
$N_{\rm sub} = 60$ is realized at the lower boundary of the cyan band. 
The green and blue dotted lines in Fig.~\ref{fig:MissingSat}
respectively indicate where $N_{\rm sub} = 60$ is realized for $M_{\rm
halo} = 10^{12} M_{\odot} / h$ and $M_{\rm halo} = 3 \cdot 10^{12}
M_{\odot} / h$;
these values for the MW halo mass roughly correspond to the current
observational limits~\cite{Cautun:2014dda,Wang:2015ala}.
For $M_{\rm halo} = 10^{12} M_{\odot} / h$, even with the pure CDM case the
satellite number is as low as $N_{\rm sub} = 94$, indicating that
the {\emph{missing satellite}} problem
itself is ameliorated if the MW halo mass takes a value close to its lower bound.
Consequently, the green line lies inside the region allowed by the
Lyman-$\alpha$ forest.
On the other hand, a larger MW halo mass makes the problem worse
($N_{\rm sub} = 274$ for 
$M_{\rm halo} = 3 \cdot 10^{12} M_{\odot} / h$ with pure CDM),
and thus further reduces the scalar DM parameter space for
satisfying the Lyman-$\alpha$ forest constraint 
and solving the {\emph{missing satellite}} problem at the same time.
To summarize,
unless the MW halo mass is close to its current lower bound and thus the
satellite number is suppressed,
the Lyman-$\alpha$ constraint leaves very little room for scalar DM
to serve as a solution to the {\emph{missing satellite}} problem.

\section{Summary and Conclusions}
\label{sec:conc}

Light scalars, if present in theories beyond the Standard Model, are
inevitably produced in the early universe due to a vacuum misalignment,
unless the initial conditions are fine tuned. Since their
interactions with other fields are typically suppressed, such light scalars
would survive until the present universe
and constitute a fraction of DM.
The goal of this work was to investigate general constraints on theories
that contain ultralight scalar fields, by analyzing the imprint of scalar DM
on the Lyman-$\alpha$ forest and studying its cosmological consequences.
The basic assumptions we made about the scalar is that it is
light, long-lived, and mainly produced by a vacuum misalignment.
If there are additional processes that produce the scalars, then the
scalar DM density would increase and the constraints would become more
stringent; in this sense our bounds are conservative.

\begin{figure}[t]
  \begin{center}
  \begin{center}
  \includegraphics[width=0.7\linewidth]{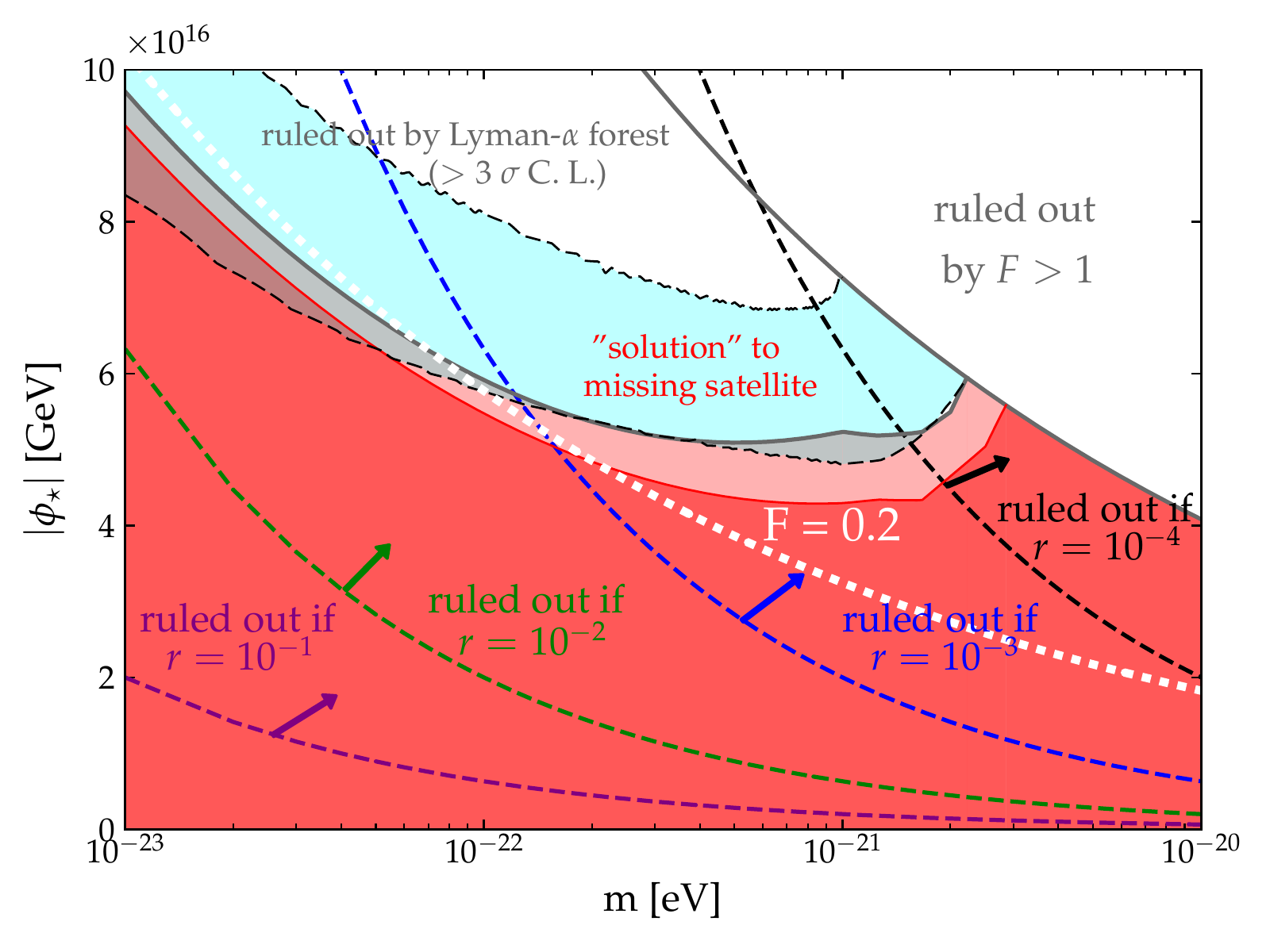}
  \end{center}
   \caption{\emph{Summary of constraints on the scalar field mass~$m$
   and the initial displacement~$\phi_\star$.
   (For an axion-like field, $\phi_\star = f_a \theta_\star$.)
   The 2 and 3$\sigma$ C.L. regions allowed by Lyman-$\alpha$ forest data are shown in
   red. The upper-right corner is excluded by the overabundance of
   DM. The dashed lines indicate the parameter regions that will be
   ruled out by a detection of a tensor-to-scalar ratio~$r$. The cyan
   band bounded by dashed lines shows where the missing satellite
   problem can be solved. On the white dotted contour, scalar DM
   constitutes 20\% of the total DM.}} 
  \label{fig:summary}
  \end{center}
\end{figure}

The results of this paper are summarized in Figure~\ref{fig:summary},
which shows the allowed values for the mass~$m$ and
initial displacement of the scalar field~$\phi_\star$.
The field displacement is generically bounded as
$\abs{\phi_\star} \lesssim 10^{16}\, \mathrm{GeV}$;
otherwise the scalar would either lead to too much DM in the universe,
or suppress structure formation and contradict
the Lyman-$\alpha$ forest measurements.
(If the scalar is an axion-like field, the bound on $\phi_\star$
corresponds to that on the product of the axion decay constant and
the initial misalignment angle, $f_a \theta_\star$, when anharmonic effects
are negligible.) 
By combining the Lyman-$\alpha$ constraints with the CMB bounds on DM
isocurvature perturbations, we further derived upper limits on the
scale of cosmic inflation in the presence of scalar DM.
These are shown in in Figure~\ref{fig:summary} as dashed lines,
indicating the parameter 
regions that will be ruled out if primordial gravitational waves are
detected in the future.
A dotted white line is also overlaid to indicate where the fraction of
the DM in scalar DM is~20\%;
this value serves as the fraction threshold below which the
Lyman-$\alpha$ forest becomes insensitive to the presence of scalar DM.

We also estimated how well scalar DM can solve the
``small-scale crisis'' of CDM.
The cyan band bounded by dashed lines in the figure
corresponds to that shown in Figure~\ref{fig:MissingSat},
indicating the
parameter region where the 
missing satellite problem is solved without the aid of baryonic physics.
With the tiny overlap between the solving region and the allowed window,
our analyses suggest that ultralight scalar DM cannot solve the 
missing satellite problem without spoiling the Lyman-$\alpha$ forest.
However we should also remark that we have used
rather simple analytic approximations for estimating the satellite
number, hence it would be important to verify this conclusion with
numerical simulations.

In this paper we have discussed cosmological implications of light scalars
that follow from the Lyman-$\alpha$ constraints with minimal
assumptions about the scalar field theory.
Thus, theories with light scalars in general are subject to our constraints.
We  focused in particular on gravitational effects, without making
assumptions about the coupling of the scalar to other 
matter fields
(except  that the couplings are small enough so that the scalar DM
survives until today).
Other possible gravitational consequences of light scalars we did not
discuss include superradiance of black
holes~\cite{Arvanitaki:2009fg,Arvanitaki:2010sy}, 
and effects on pulsar timing observations~\cite{Khmelnitsky:2013lxt}
or binary pulsars~\cite{Blas:2016ddr}.
We also remark that concrete models of scalar DM can contain
couplings with other fields, such as axion-type couplings to photons.
In such cases the model parameters can further be constrained from
various experiments~\cite{Ringwald:2012hr}.
It would also be interesting to combine such coupling constraints with the
results of this paper, to systematically analyze
specific classes of scalar DM models.

\section*{Acknowledgments}

We thank Daniel Grin, Wayne Hu, Serguey Petcov, Vikram Rentala, Piero
Ullio, and Mauro Valli for helpful conversations. 
T.K., M.V., and R.M. acknowledge support from the INFN INDARK PD51 grant. 
M.V. also acknowledges support from the ERC-StG cosmoIGM.
V.I. is supported by US NSF grant AST-1514734.


\appendix

\section{Exact Solution of the Klein--Gordon Equation}
\label{app:onset}

In this appendix we provide the exact solution of the
homogeneous Klein--Gordon equation~(\ref{homoKG}) in a
radiation-dominated universe whose Hubble rate redshifts as
\begin{equation}
 H \propto a^{-2}.
  \label{A.1}
\end{equation}
Here we should note that the redshifting of the Hubble rate can depart
from $\propto a^{-2}$ 
when the effective number of relativistic degrees of freedom~$g_{(s)*}$
changes in time;
however as long as $g_{(s)*}$ stays constant while
the scalar makes the transition from vacuum energy-like
(non-oscillatory) to matter-like (oscillatory),
then the solution under~(\ref{A.1}) can be used to
accurately compute the scalar density in the
asymptotic future. 

The solution of (\ref{homoKG}) with (\ref{A.1}) is given in terms of 
a Bessel function of the first kind as
\begin{equation}
 \phi = \phi_{\star} \, \Gamma \left( \frac{5}{4}\right)
  \left( \frac{4 H}{m} \right)^{1/4}
  J_{1/4} \left(\frac{m}{2 H}\right),
\end{equation}
where we have chosen the initial condition for $\phi$ such that
it approaches a constant value
$\phi \to \phi_{\star}$ as $m / H \to 0$.
Hence, after the scalar starts to oscillate, 
its density~(\ref{rhoandp}) asymptotes to
\begin{equation}
 \lim_{ \frac{m}{H} \to \infty } \rho_\phi =
  \frac{4}{\pi }
  \left[ \Gamma \left(\frac{5}{4}\right) \right]^2
  m^{1/2} \phi_{\star}^2 \, H^{3/2}  .
\end{equation}
Equating this with (\ref{rho_late}):
\begin{equation}
 \lim_{ \frac{m}{H} \to \infty } \rho_\phi = 
  \frac{1}{2} m^2 \phi_{\star}^2
  \left(\frac{a_{\mathrm{osc}}}{a}\right)^3
  =   \frac{1}{2} m^2 \phi_{\star}^2
  \left(\frac{H}{H_{\mathrm{osc}}}\right)^{3/2} ,
\end{equation}
yields the ratio between the scalar mass and
$H_{\mathrm{osc}}$ as
\begin{equation}
 \frac{m^2}{H_{\mathrm{osc}}^2} = 
  \left( \frac{8}{\pi }  \right)^{4/3}
  \left[\Gamma \left(\frac{5}{4} \right) \right]^{8/3}
  \approx 2.68.
\end{equation}
With this $H_{\mathrm{osc}}$,
one can compute the scalar density not only during radiation-domination, but
also in the subsequent epochs
as demonstrated in Section~\ref{subsec:abundance}.

\section{Linear Matter Power Spectrum with Scalar DM}
\label{app:lmps}

Here we compute the linear matter power spectrum in the presence of scalar DM.
The goal of this appendix is to clarify the effects of the scalar DM
parameters on the linear matter power spectrum through 
(mostly) analytic computations. 

\subsection{Scalar DM as a Fluid}
\label{subsec:fluid}

We start by describing scalar DM as a cosmological fluid,
using the formalism discussed in,
e.g.,~\cite{Arvanitaki:2009fg,Hu:2000ke,Hui:2016ltb,Widrow:1993qq}.
A large number of scalar particles with mass~$m$ behaves as a
classical field obeying the Klein--Gordon and Einstein's equations,
\begin{equation}
 \nabla_\mu \nabla^\mu \phi = m^2 \phi,
  \quad
  G_{\mu \nu} = 8 \pi G \, T_{\mu \nu},
  \label{KG}
\end{equation}
where the scalar field contributes to the energy-momentum tensor as
\begin{equation}
 T_{\mu \nu}^\phi = g_{\mu \nu}
  \left( -\frac{1}{2} \partial_\rho \phi \partial^\rho \phi -
   \frac{1}{2} m^2 \phi^2 \right)
  + \partial_\mu \phi \partial_\nu \phi.
\end{equation}
Regarding the metric of the FRW universe with perturbations,
we take the Newtonian gauge and ignore anisotropic stress, 
\begin{equation}
 ds^2 = - (1 + 2 \Phi) dt^2 + a(t)^2 (1- 2 \Phi) d \bd{x}^2. 
\end{equation}
The Hubble rate is defined as $H = \dot{a} / a$, with an overdot
denoting a $t$-derivative. 

When $H \ll m$, and hence the scalar field is harmonically oscillating,
it is convenient to rewrite $\phi$ as
\begin{equation}
 \phi = \frac{1}{\sqrt{2 m}}
  \left( \varphi e^{-imt} + \varphi^* e^{imt} \right),
\end{equation}
in terms of a complex field~$\varphi$ describing the oscillation
amplitude whose time dependence is slow compared to the oscillation period.
We rewrite the Klein--Gordon equation in terms of~$\varphi$
under the assumption of tiny perturbations $\abs{\Phi} \ll 1$,
and thus ignoring terms that 
contain quadratic or higher orders of~$\Phi$.
Furthermore, let us focus on nonrelativistic modes ($k / a \ll m$),
and suppose the time scales for the variations of 
$\Phi$, $\varphi$, $a$, and their derivatives to be much longer than the
oscillation period,
i.e., $\abs{\dot{\Phi}} \ll m \abs{\Phi}$, $\abs{\ddot{\varphi}} \ll m
\abs{\dot{\varphi}}$, etc.
This, in particular,  allows us to drop second time-derivatives
in the equation.
After taking an average over the oscillation period,
the Klein--Gordon equation reduces to a Schr{\"o}dinger-type
equation~\cite{Widrow:1993qq}, 
\begin{equation}
 i\left(\dot{\varphi}  + \frac{3}{2} H \varphi \right) =
  -\frac{\partial^2 \varphi}{2 a^2 m}  +
  m \Phi \varphi,
\label{schr}
\end{equation}
where $\partial^2 \equiv \partial_i \partial_i$, and sum over repeated
spatial indices is implied irrespective of their positions.

In terms of the amplitude and the phase of~$\varphi$, we now define
\begin{equation}
 \rho_{\phi} \equiv
m \varphi \varphi^*,
  \quad
  v_i \equiv
  \frac{\partial_i \{ \arg (\varphi) \} }{a m }
  = 
- \frac{i}{ 2 a m}
  \left( \frac{\partial_i \varphi}{\varphi } - \frac{\partial_i
   \varphi^*}{\varphi^*}  \right),
\end{equation}
whose meaning will soon become clear.
Multiplying both sides of the Schr{\"o}dinger equation~(\ref{schr})
by~$\varphi^*$, its real and imaginary parts
respectively lead to the following equations, 
\begin{gather}
 \dot{v}_i + H v_i + \frac{v_j \partial_j v_i}{a}
  = -\frac{\partial_i \Phi}{a} +
  \frac{1}{2 a^3 m^2} \partial_i \left( \frac{\partial^2
				  \sqrt{\rho_{\phi}} }{\sqrt{\rho_{\phi} }} \right),
  \label{maru-vi}
 \\
 \dot{\rho}_{\phi} + 3 H \rho_{\phi} + \frac{\partial_i (\rho_{\phi} v_i)}{a} = 0.
\label{maru-v}
\end{gather}
One can also rewrite the Einstein's equation in a similar fashion.
Focusing on sub-horizon and nonrelativistic modes,
i.e. $ H \ll k/a \ll m$,
it can be checked that the $(0,0)$~component of the Einstein's equation yields
\begin{equation}
 \frac{\partial^2 \Phi }{a^2} =
  4 \pi G \left( \rho_{\phi} + T_{00}^{\mathrm{others}} \right)
  - \frac{3}{2} H^2,
  \label{maru-vii}
\end{equation}
where $T_{00}^{\mathrm{others}}$ denotes the contributions to the
energy-momentum tensor from components other than the scalar DM.
Interpreting $\rho_\theta$ and $v_i$ as the density and velocity
fields, the set of equations
(\ref{maru-vi}), (\ref{maru-v}), and (\ref{maru-vii}) are seen to
correspond respectively to the Euler, continuity, and Poisson equations;
thus we have arrived at a fluid description of the scalar DM.
The only difference with the familiar CDM fluid is the existence of the
last term in the right hand side of~(\ref{maru-vi}), 
which represents a pressure due to the wave nature of the scalar field
on small scales.

\subsection{Evolution of DM Density Perturbations}

Let us now study the evolution of density perturbations in a
matter-dominated universe that is filled with the scalar DM and CDM.
We also describe CDM
as a fluid obeying the 
Euler and continuity equations as in (\ref{maru-vi}) and
(\ref{maru-v}), except for that there is no pressure term,
and replace $T_{00}^{\mathrm{others}}$ in the Poisson
equation~(\ref{maru-vii}) by the CDM density~$\rho_{\mathrm{c}}$.
Then one immediately sees from the equations that the homogeneous and
isotropic background satisfies
\begin{equation}
 \frac{\dot{\bar{\rho}}_\phi}{\bar{\rho}_\phi}
  = \frac{\dot{\bar{\rho}}_\mathrm{c}}{\bar{\rho}_\mathrm{c}} = -3 H ,
  \quad
  \bar{\rho}_\phi + \bar{\rho}_{\mathrm{c}} =  \frac{3 H^2}{8 \pi G} ,
\label{bg_MD}
\end{equation}
where a bar is used to denote unperturbed values.
We discuss the density fluctuations around the background in
terms of the density contrast,
\begin{equation}
 \delta_n = \frac{\rho_n - \bar{\rho}_n}{\bar{\rho}_n},
\end{equation}
where $n = \phi, \mathrm{c}, \mathrm{m}$, with
$\rho_{\mathrm{m}} = \rho_\phi + \rho_{\mathrm{c}}$ being the total DM density.
Expressing the ratio between the unperturbed densities of the scalar and
total DM as 
\begin{equation}
 F = \frac{\bar{\rho}_\phi }{\bar{\rho}_\mathrm{m}},
\end{equation}
which is a constant in the range $ 0 \leq F \leq 1$,
the density contrast of the total DM is written as
\begin{equation}
 \delta_\mathrm{m} = F \delta_\phi + (1- F) \delta_{\mathrm{c}}. 
\end{equation}
Expanding the equations (\ref{maru-vi}), (\ref{maru-v}), and
(\ref{maru-vii}) up to linear order in $\delta$ and $v_{i}$,
and then combining the equations to 
eliminate $v_i$ and $\Phi$, one arrives at the evolution equations for
the density contrasts,
\begin{gather}
 \ddot{\delta}_{\phi \bd{k}} + 2 H \dot{\delta}_{\phi \bd{k}}
  + \frac{c_s^2 k^2}{a^2} \delta_{\phi \bd{k}}
 - \frac{3}{2} H^2 \delta_{\mathrm{m}\bd{k}} = 0,
 \label{delta-phi}
\\
 \ddot{\delta}_{\mathrm{c} \bd{k}} + 2 H \dot{\delta}_{\mathrm{c} \bd{k}}
 - \frac{3}{2} H^2 \delta_{\mathrm{m}\bd{k}} = 0.
 \label{delta-c}
\end{gather}
Here we have written the linearized equations in terms of the Fourier
components ($\bd{k}$ is a comoving wave number, with $k = \abs{\bd{k}}$),
and the sound speed of the scalar DM is  
\begin{equation}
 c_s^2 \equiv \frac{k^2}{4 a^2 m^2}.
\end{equation}
From their derivations, it should be noted that 
the equations (\ref{delta-phi}) and (\ref{delta-c})
are valid during the matter-dominated epoch, and
for wave numbers that are sub-horizon and nonrelativistic,
i.e., \mbox{$H \ll k/a \ll m$}.

In the case where the scalar DM constitutes the entire DM, i.e. $F = 1$,
one can read off its Jeans wave number from the last two terms in
(\ref{delta-phi}) as
\begin{equation}
 \frac{k_{\mathrm{J}}}{a} = \sqrt{Hm},
\label{eq:Jeans}
\end{equation}
where we have ignored numerical factors.
As we are interested in scalar masses larger than the Hubble rate
at matter-radiation equality, the Jeans length is smaller than the
Hubble length throughout the matter-dominated epoch.
On length scales larger than the Jeans length ($k < k_{\mathrm{J}}$),
the pressure term is negligible and the scalar DM behaves similarly to CDM;
thus the density fluctuation possesses the usual growing mode 
$\delta_{\phi \bd{k}} \propto a$.
However below the Jeans length ($k > k_{\mathrm{J}}$),
$\delta_{\phi \bd{k}}$ undergoes oscillations of
$ \propto \exp (\pm 2 i c_s k / a H)$
and thus does not grow.
Here note that in a matter-dominated universe, 
the comoving Jeans wave number grows as $k_{\mathrm{J}} \propto a^{1/4}$.
Hence for modes that are sub-Jeans at the time of
matter-radiation equality, i.e. $k > k_{\mathrm{Jeq}}$,
the scalar DM perturbation is prevented from growing until the mode~$k$
crosses the Jeans scale at $a = a_k$, where
\begin{equation}
 a_k = a_{\mathrm{eq}} \left( \frac{k}{k_{\mathrm{J eq}}} \right)^{4}.
\label{a_k}
\end{equation}
With a slight abuse of language, we will refer to $k_\mathrm{J}$ as
defined in~(\ref{eq:Jeans}) as the Jeans wave number even in
cases with $F \neq 1$.

If DM is a mixture of scalar DM and CDM, i.e. $0 <
F < 1$, then the matter fluctuations grow even on small scales,
albeit slowly.
This is seen by dropping $\delta_{\phi \bd{k}}$
in~(\ref{delta-c}) by
taking an average over $\delta_{\phi \bd{k}}$'s oscillation period, giving
\begin{equation}
  \frac{d^2 \delta_{\mathrm{c} \bd{k}} }{d a^2} +
 \frac{3}{2a }\frac{d \delta_{\mathrm{c} \bd{k}} }{d a}
 - \frac{3 (1-F) \delta_{\mathrm{c}\bd{k}}}{2 a^2 } = 0.
\end{equation}
Here we have rewritten the derivatives in terms of time by those of~$a$,
using $H \propto a^{-3/2}$ (cf.~(\ref{bg_MD})).
This equation has a general solution of
\begin{equation}
 \delta_{\mathrm{c} \bd{k}} = C_+ a^{n_+} + C_- a^{n_-},
  \quad
  \mathrm{with}
  \quad
  n_{\pm} = \frac{-1 \pm \sqrt{25 - 24 F} }{4},
  \label{n_pm}
\end{equation}
whose first term is a growing mode, and the second a decaying mode.
The growing CDM perturbation drags the scalar DM and thus $\delta_{\phi
\bd{k}}$ starts to grow even before crossing the Jeans
scale~(\ref{eq:Jeans}).

On the other hand, wave modes with $ k < k_{\mathrm{Jeq}}$ stay super-Jeans
throughout the matter-dominated epoch.
We should also note that during radiation-domination, there is no
significant growth for both scalar DM and CDM perturbations on
sub-horizon scales.\footnote{The evolution of 
the matter fluctuations prior to matter-radiation equality
can be described by further including a homogeneous radiation component
$\bar{\rho}_\mathrm{r} \propto a^{-4}$
to~$T_{00}^{\mathrm{others}}$ in~(\ref{maru-vii})
(hence ignoring radiation perturbations on sub-horizon scales);
this amounts to multiplying
the term $(3/2) H^2 \delta_{\mathrm{m}\bd{k}}$
in (\ref{delta-phi}) and (\ref{delta-c}) by 
$\bar{\rho}_\mathrm{m} / ( \bar{\rho}_\mathrm{m} +
\bar{\rho}_\mathrm{r})$. During radiation-domination
($\bar{\rho}_\mathrm{m} / \bar{\rho}_\mathrm{r} \to 0$),
$\delta_{\phi \bd{k}} $ has general solutions of 
$\exp \{\pm i (c_s k / a H) \ln a \} $.\label{foot:RD}}
Therefore the difference in the matter perturbations between cases
with and without scalar DM becomes prominent
on wave numbers $k > k_{\mathrm{Jeq}}$,
mainly due to the difference in the evolution during the matter-dominated epoch.
For reference, the Jeans wave number at the matter-radiation equality is
\begin{equation}
 \frac{k_{\mathrm{J eq}}}{a_0}
=  \frac{a_\mathrm{eq}}{a_0}  \sqrt{H_{\mathrm{eq}} m}
\approx 7  \, \mathrm{Mpc}^{\text{-}1}
  \left( \frac{m}{10^{-22} \, \mathrm{eV}} \right)^{1/2},
  \label{kJ_eq}
\end{equation}
where the subscript~``$0$'' denotes quantities today.

\subsection{Suppression of Linear Matter Power Spectrum}
\label{ap:suppression}

In order to evaluate the growth of the perturbations for arbitrary~$F$,
we suppose an adiabatic initial condition and consider
$\delta_\phi$ to behave similarly to $\delta_{\mathrm{c}}$
during radiation-domination, and on super-Jeans scales.
For wave numbers $ k > k_{\mathrm{Jeq}}$, let us make a rough approximation that 
the total matter perturbation follows $\delta_{\mathrm{m}\bd{k}} \propto a^{n_+}$
since matter-radiation equality until crossing the Jeans scale at $a = a_k$,
then subsequently grows as the usual $\delta_{\mathrm{m}\bd{k}} \propto a$.
This approximation is crude, but allows us to understand the overall spectral
shape of the suppression and how it is determined by the scalar DM
parameters.

By comparison with the pure CDM ($F = 0$) case where
$\delta_{\mathrm{m}\bd{k}} \propto a$ throughout 
matter-domination, one can estimate the suppression of the linear matter
power spectrum due to the scalar DM. 
For wave modes that have crossed the Jeans scale, the matter perturbation
containing scalar DM is suppressed relative to that with pure CDM by
\begin{equation}
 \left|
  \frac{\delta_{\mathrm{m}\bd{k}}^{(\phi + \mathrm{c})}}
  {\delta_{\mathrm{m}\bd{k}}^{(\mathrm{c})}}
  \right|_{k_{\mathrm{Jeq}} < k < k_{\mathrm{J}}}
=
 \left(\frac{a_{\mathrm{eq}}}{a_k}\right)^{1-n_+} =
 \left(\frac{k_{\mathrm{Jeq}}}{k}\right)^{4 (1-n_+)}.
\label{supp-20}
\end{equation}
On the other hand for modes that are still sub-Jeans at the time the
matter perturbations are measured, the suppression saturates to a 
$k$-independent value of
\begin{equation}
 \left|
  \frac{\delta_{\mathrm{m}\bd{k}}^{(\phi + \mathrm{c})}}
  {\delta_{\mathrm{m}\bd{k}}^{(\mathrm{c})}}
  \right|_{k > k_{\mathrm{J}}}
 =
 \left(\frac{a_{\mathrm{eq}}}{a}\right)^{1-n_+} =
 \left(\frac{k_{\mathrm{Jeq}}}{k_{\mathrm{J}} } \right)^{4 (1-n_+)}.
 \label{supp-21}
\end{equation}
See also \cite{Arvanitaki:2009fg} where similar results were obtained.

\begin{figure}[t]
 \begin{minipage}{.53\linewidth}
  \begin{center}
 \includegraphics[width=\linewidth]{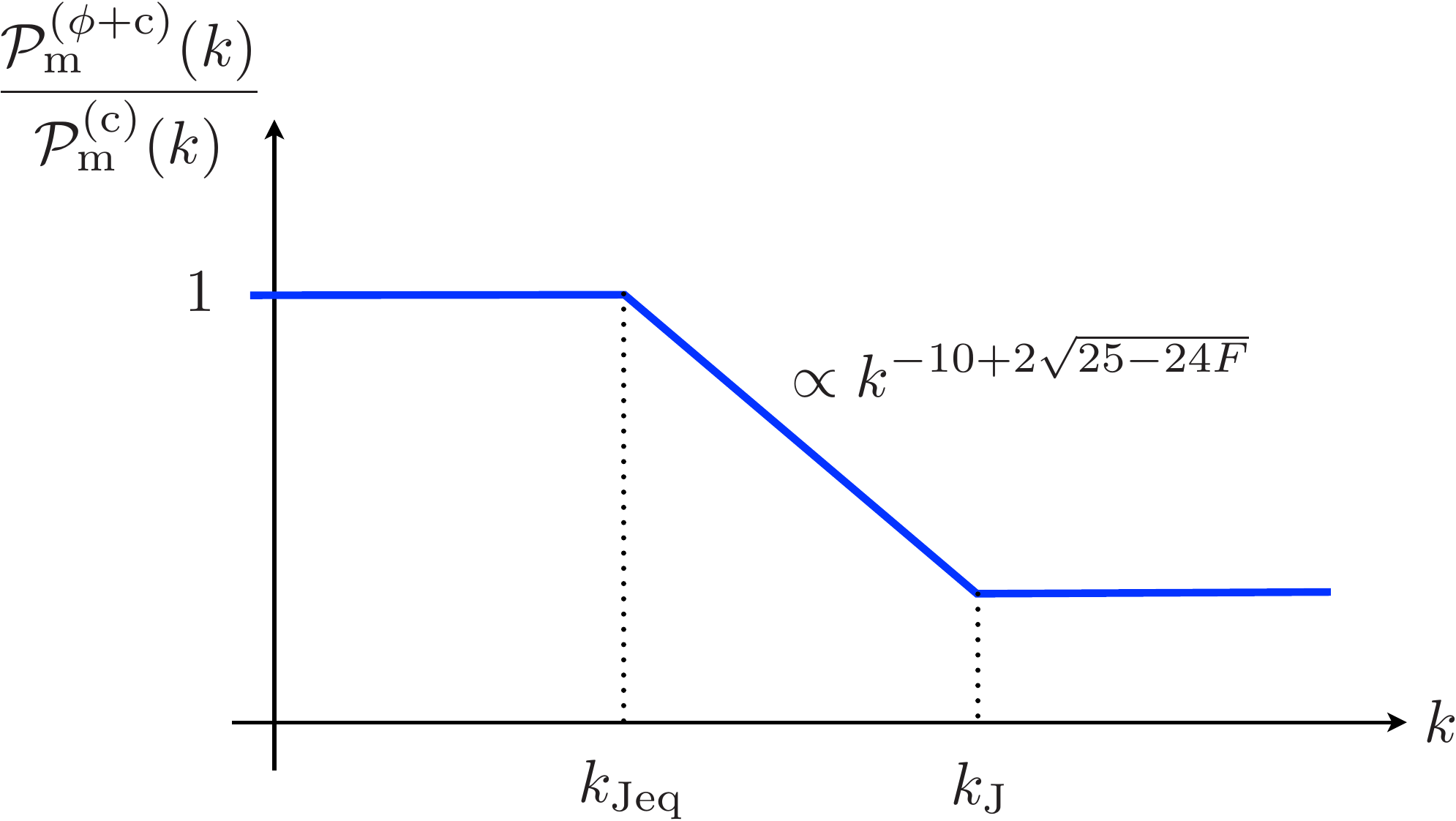}
  \end{center}
 \end{minipage} 
 \begin{minipage}{0.01\linewidth} 
  \begin{center}
  \end{center}
 \end{minipage} 
 \begin{minipage}{.40\linewidth}
  \begin{center}
 \includegraphics[width=\linewidth]{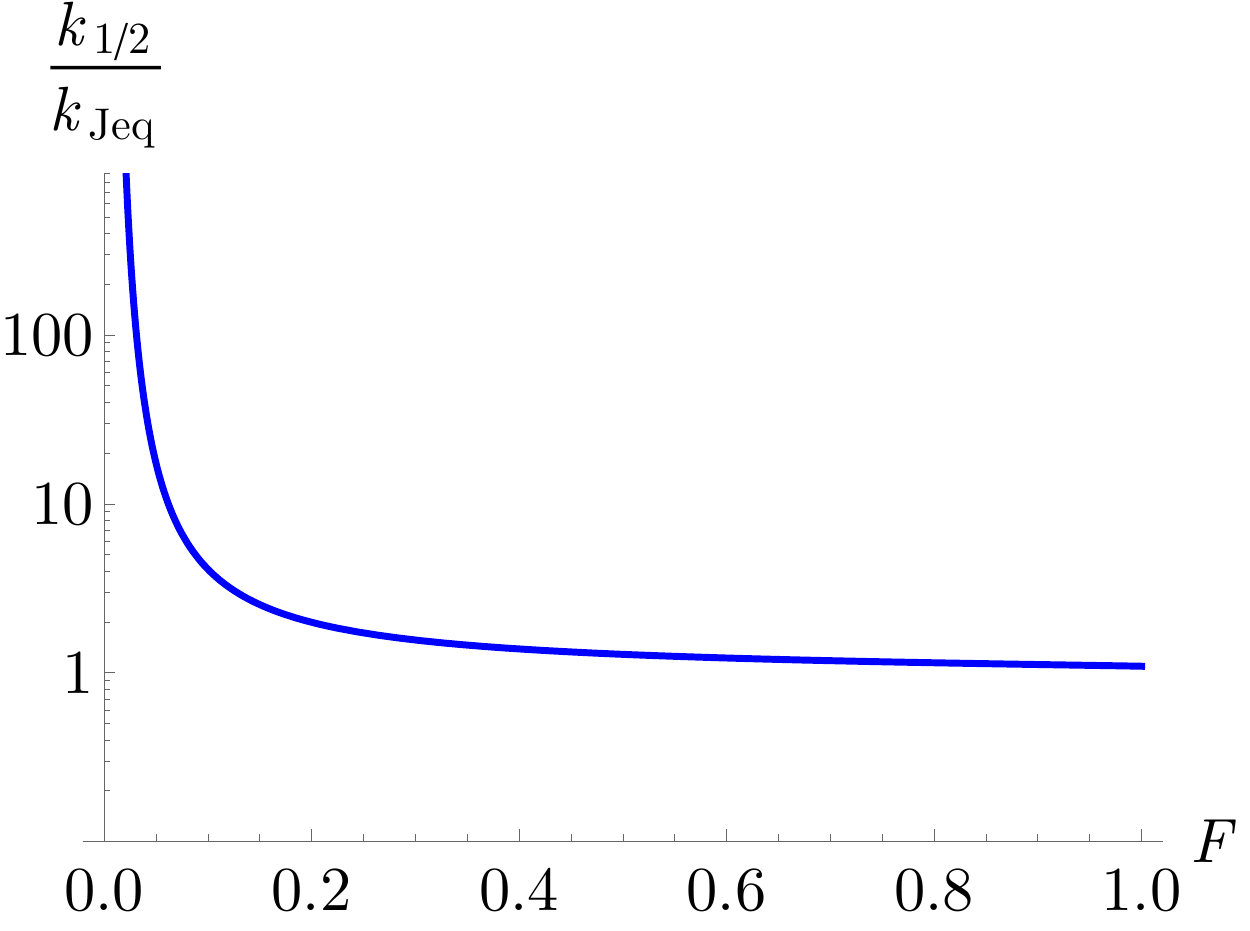}
  \end{center}
 \end{minipage} 
 \caption{\emph{Left: A sketch of the suppression of the linear matter power
 spectrum containing scalar DM relative to that with pure CDM.
 Right: The wave number~$k_{1/2}$ at which the linear matter
 power spectrum is suppressed by~$1/2$, in units of the Jeans wave number at
 matter-radiation equality~$k_{\mathrm{Jeq}}$, as a function of the
 scalar DM fraction~$F$.}}
 \label{fig:supp-ana}
\end{figure}

The relative suppression of the linear matter power spectrum
$\mathcal{P}_{\mathrm{m}} (k) \propto \abs{\delta_{\mathrm{m} \bd{k}}}^2$
due to the scalar DM, obtained from squaring (\ref{supp-20}) and (\ref{supp-21}), 
is sketched in the left panel of Figure~\ref{fig:supp-ana}.
The actual spectrum with scalar DM can be oscillatory, and so what is
illustrated in the figure should be considered as the envelope.
We stress that while the scalar mass~$m$ determines the
Jeans wave number at equality~$k_{\mathrm{Jeq}}$ above which the 
suppression appears, the spectral index of the suppression is set by the
fraction~$F$.
Moreover, the suppression saturates at the Jeans wave
number~$k_{\mathrm{J}}$ at the time of the measurement,
where the saturated suppression factor~(\ref{supp-21}) is independent
of~$m$.\footnote{The suppression function presented in Eq.~(8)
of~\cite{Hu:2000ke} behaves quite differently from what we 
discussed in the region $ k \gg k_{\mathrm{Jeq}}$. 
This is because the function of~\cite{Hu:2000ke} was obtained to
describe the first few efolds of suppression at around~$k_{\mathrm{Jeq}}$, instead of
the asymptotic behavior.
We thank Wayne Hu for private communication on this point.}

Provided that the square of~(\ref{supp-21}) is smaller than $1/2$, we
can define the wave 
number~$k_{1/2}$ at which the power spectrum~$\mathcal{P}_{\mathrm{m}} (k) $
is suppressed by $1/2$.
From~(\ref{supp-20}), $k_{1/2}$ is obtained
in terms of the Jeans scale at equality~(\ref{kJ_eq}) as
\begin{equation}
 \frac{k_{1/2}}{k_{\mathrm{Jeq}}} = f(F),
  \quad
  \mathrm{where}
  \quad
  f(F) = 2^{\frac{1}{10- 2\sqrt{25-24 F }}}.
  \label{k0.5analytic}
\end{equation}
This ratio as a function of~$F$ describes how the suppression effect
is diluted for a smaller scalar DM fraction.
We plot this in the right panel of
Figure~\ref{fig:supp-ana}, where one sees that 
$k_{1/2} \sim  k_{\mathrm{Jeq}}$ for $F > 0.1$,
whereas $k_{1/2} $ becomes exponentially larger than $ k_{\mathrm{Jeq}}$
for $F \lesssim 0.1$.
In particular if $F \lesssim 0.07$, 
then $k_{1/2}$ even exceeds the present-day Jeans scale
$k_{\mathrm{J}0} = (a0/a_{\mathrm{eq}})^{1/4} k_{\mathrm{Jeq}} \sim
(3000)^{1/4} k_{\mathrm{Jeq}}$
(here we are ignoring dark energy for simplicity);
namely, the linear power spectrum today does not fall below $1/2$ of
that from pure CDM.
This indicates that a scalar DM, no matter how light its mass is,
would not impact structure formation
as long as its fraction is below~$\sim 10\%$ of the total DM.

\begin{figure}[t]
  \begin{center}
  \begin{center}
  \includegraphics[width=0.7\linewidth]{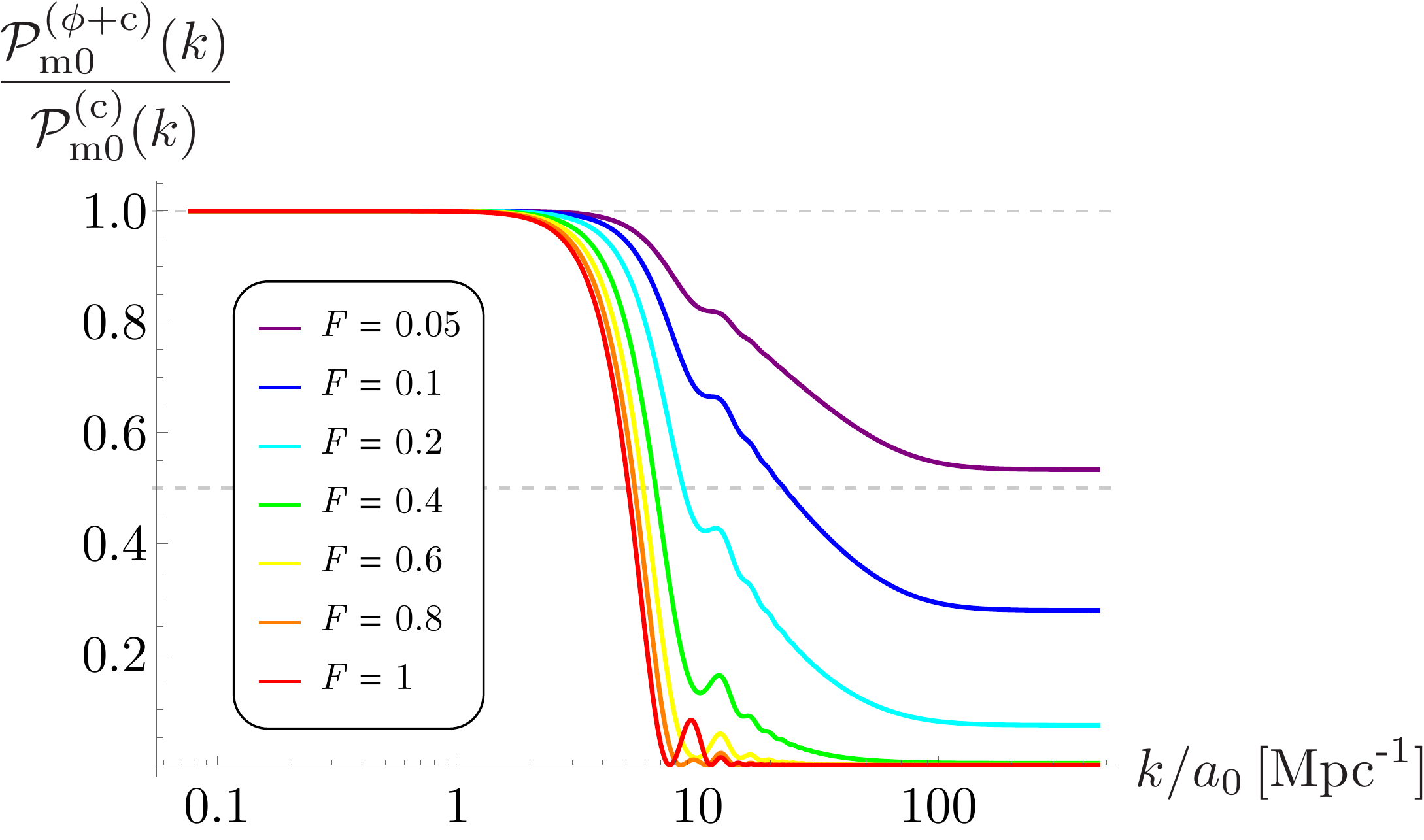}
  \end{center}
   \caption{\emph{Suppression of the linear matter power spectrum today due to
   a scalar DM with mass $m = 10^{-22}\, \mathrm{eV}$, for various
   values of the scalar DM fraction~$F$. The results are obtained by
   numerically solving the evolution equations (\ref{delta-phi}) and (\ref{delta-c}).}}
  \label{fig:supp-num}
  \end{center}
\end{figure}

We have also numerically solved the coupled evolution equations
(\ref{delta-phi}) and (\ref{delta-c}).
In order to incorporate the slow growth of the CDM
perturbation towards the end of the radiation-dominated epoch, 
we further included a background radiation component as explained in
Footnote~\ref{foot:RD},
and started the computations at $a = a_{\mathrm{eq}} / 10$
with a simplified adiabatic initial condition
$\delta_{\phi \bd{k}} = \delta_{\mathrm{c} \bd{k}}$,
$\dot{\delta}_{\phi \bd{k}} = \dot{\delta}_{\mathrm{c} \bd{k}} =
0$.\footnote{A more rigorous treatment would involve solving the 
relativistic perturbation equations until the fluid description becomes
valid (see
e.g. \cite{Khlopov:1985jw,Nambu:1989kh,Ratra:1990me,Hwang:1996xd}), and
also inclusion of other components such baryons and dark energy. 
However our simplified treatment should suffice for estimating
the relative suppression of the perturbations by the scalar DM at the
order-of-magnitude level.}
The ratio of the resulting linear matter power spectrum today between cases
with scalar DM and pure CDM is shown in Figure~\ref{fig:supp-num},
for various values of the scalar DM fraction~$F$. 
The mass of the scalar DM is fixed to $m = 10^{-22}\, \mathrm{eV}$,
and the spectrum is shown for wave numbers that are sub-horizon and
nonrelativistic ($H < k/a < m$) at the initial time of the calculation.
As was indicated by the analytic arguments, 
the suppression appears at around the Jeans scale at equality
$k_{\mathrm{Jeq}}/a_0 \sim 7 \, \mathrm{Mpc}^{\text{-}1}$,
and saturates at around the Jeans scale today
$k_{\mathrm{J} 0}/a_0 \sim 50 \, \mathrm{Mpc}^{\text{-}1}$.
It is also clearly seen that 
$k_{1/2} \sim k_{\mathrm{Jeq}}$ for $F > 0.1$,
whereas for $F \lesssim 0.1$ the suppression does not fall much
below~$50\%$ on any scale.

The scale~$k_{1/2}$ also offers a rough guide to estimate the mass of halos
whose formation are suppressed.
The mass contained in a sphere of diameter $2 \pi a / k_{1/2}$ is
\begin{equation}
 M_{1/2} = \frac{H^2}{2 G} \left( \frac{\pi a}{k_{1/2}} \right)^3
 \sim 10^{10} M_{\odot} \, f(F)^{-3}
  \left( \frac{m}{10^{-22} \, \mathrm{eV}} \right)^{-3/2},
\end{equation}
where in the far right hand side we
substituted~(\ref{k0.5analytic}) to~$k_{1/2}$,
and used that $M_{1/2}$ is time-independent during matter-domination and
thus estimated its value at equality.
We can thus infer from linear theory
that the scalar DM suppresses the number of halos with masses below~$M_{1/2}$.

\section{The Area Criterion for the Lyman-$\alpha$ Forest}
\label{ap:area}

\begin{figure}[t]
  \begin{center}
  \begin{center}
  \includegraphics[width=0.7\linewidth]{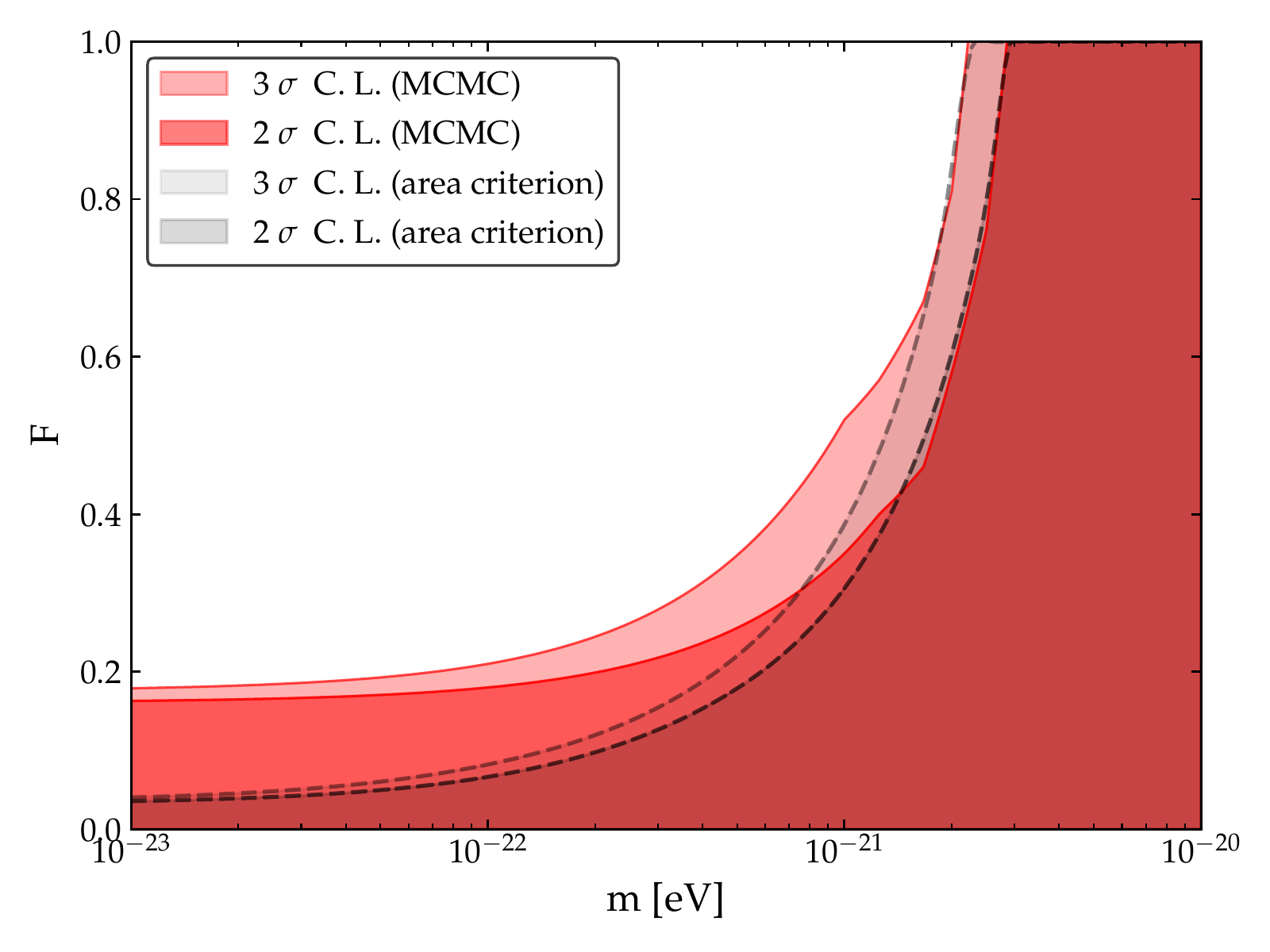}
  \end{center}
   \caption{\emph{Here we compare the 2 and 3~$\sigma$ limits from the
   MCMC analysis (red regions), with the 2 and 3~$\sigma$ limits
   determined through the area criterion (superposed gray regions
   bounded by dashed lines). 
   All the combinations of scalar DM mass and fraction which sample the region outside the gray contours are thus rejected by the area criterion analysis.}}
  \label{fig:areaVSmcmc}
  \end{center}
\end{figure}

\looseness=-1 The aim of this Appendix is to compare the Lyman-$\alpha$ forest constraints determined through the full MCMC analysis discussed in detail in Section~\ref{sec:lyman}, against the bounds obtained by applying a simple yet intuitive method introduced by the authors of Ref.~\cite{Murgia:2017lwo}, i.e.~the {\emph{area criterion}} for the Lyman-$\alpha$ forest.

Whereas absolute limits on DM properties can only be extracted through a comprehensive statistical analysis as the one that we have performed, the area criterion allows to investigate deviations with respect to a given reference model in an approximate yet immediate way, without the need of running computationally expensive cosmological simulations.
Firstly, we parameterize the deviation of a given model with respect to the standard CDM case through the ratio
\begin{equation}\label{eq:rk}
 \xi(k) = \frac{P_{1\rm{D}}(k)}{P^{\rm{CDM}}_{1\rm{D}}(k)},
\end{equation}
where $P_{1\rm{D}}(k)$ is the 1D matter power spectrum of the model that we are considering, computed by the following integral on the 3D matter power spectrum:
\begin{equation}\label{eq:pk1d}
 P_{1\rm{D}}(k)=\frac{1}{2\pi} \int\limits_k^\infty {\rm d}k'k'P(k'),
\end{equation}
with $P(k')$ being the 3D linear matter power spectrum, computed at redshift $z=0$.

In order to find out whether a model deviates more or less from standard CDM, with respect to the reference model that we have chosen, we adopt the following criterion: a model is rejected if it shows a larger power suppression with respect to the reference one. The suppression in the power spectra is computed via the following estimator:
\begin{equation}\label{eq:deltaA}
 \delta A \equiv \frac{A_{\rm CDM} - A}{A_{\rm CDM}},
\end{equation}
where $A$ is the integral of $\xi(k)$ over the typical range of scales probed by Lyman-$\alpha$ observations (e.g.,~$0.5~h/{\rm Mpc} \leq k \leq 10~h/{\rm Mpc}$ for the MIKE/HIRES+XQ-100 combined data set used in this work and in Ref.~\cite{Irsic:2017yje}), i.e.,
\begin{equation}\label{eq:A}
A = \int\limits_{k_{\rm min}}^{k_{\rm max}} {\rm d}k\ \xi(k),
\end{equation}
such that $A_{\rm CDM} \equiv k_{\rm max} - k_{\rm min}$ by construction.

We are now able to use the area criterion~\footnote
{The physical observable for Lyman-$\alpha$ forest experiments is the
\emph{flux power spectrum} $P_{\rm{F}}(k,z)$, not the linear
matter power spectrum. Nonetheless, two different features of the
Lyman-$\alpha$ physics suggest that the area criterion analysis could be 
also quantitatively correct. Firstly, the relation between linear matter
and flux power can be modeled as $P_{\rm{F}}(k,z)=b^2(k,z)P_{1\rm{D}}^{\rm{linear}}(k,z)$, with $b^2(k,z)$ being a bias factor that differs very little for models reasonably close to the standard CDM case (see e.g.~\cite{Viel2005}): this justifies the use of Eq.~\eqref{eq:rk}. Furthermore, the area criterion is motivated by the fact that IGM peculiar velocities (typically $<100$ km/s) tend to redistribute the small-scale power within a relatively broad range of wave-numbers in the explored region~\cite{gnedin2002}.}
that we have just outlined for constraining the DM mass $m$ and abundance $F$, in the ultralight scalar DM framework that we have examined in this paper. In doing so, we calibrate the method by taking as references the 2 and 3~$\sigma$ limits on the scalar DM mass where it constitutes all the DM content of the universe, namely the values given by the intercepts between the 2 and 3~$\sigma$ contours with the $F=1$ line in Fig.~\ref{fig:m-f}.  

We have then computed the corresponding linear power spectra with the numerical Boltzmann solver {\tt axionCAMB}~\cite{Hlozek:2014lca} and plugged into Eqs.~\eqref{eq:rk} and~\eqref{eq:deltaA}. The resulting $\delta A_{\rm REF_{2\sigma}}$ and $\delta A_{\rm REF_{3\sigma}}$ are the estimate of the small-scale power suppression with respect to standard CDM for scalar DM models that are excluded at 2 and 3~$\sigma$~C.L., respectively.
Thereafter, we have built a grid in the $\{m , F\}$-space, where each grid-point is associated to a different combination of scalar DM mass and fraction, in order to compare all the corresponding $\delta A$ with $\delta A_{\rm REF}$ and accept only those combinations which exhibit a smaller power suppression, i.e.~$\delta A < \delta A_{\rm REF}$.

\looseness=-1 The results are reported in Figure~\ref{fig:areaVSmcmc},
where the red shaded areas refer to the 2 and 3~$\sigma$ contours from
the MCMC analysis, while the superposed gray areas bounded by dashed
lines correspond to the 2 and 3~$\sigma$ contours determined through the area criterion. 
\looseness=-1 Hence, all the combinations of scalar DM mass and fraction which sample the region outside the gray contours are not allowed by the area criterion analysis.

\looseness=-1 Firstly we notice that for the case $F=1$ the numbers returned by the area criterion are by definition in agreement with the more exact and comprehensive MCMC analysis discussed in Section~\ref{sec:lyman}.
\looseness=-1 However, below scalar DM abundances of around~$30\%$, the
contours predicted by applying the area criterion clearly depart from
the results of the full statistical analysis that we have
performed. This is due to the fact that when we apply the area criterion
to models with small masses, even the practically negligible
suppressions associated with small fractions (see
Appendix~\ref{ap:suppression}) correspond to larger estimators $\delta
A$ with respect to the reference one. In other words, shifting the
position of the cut-off at lower wavenumbers (i.e.~investigating small
scalar DM masses) unavoidably leads to a suppression which, quantified
through the area estimator, is larger than the reference one.
Consequently, although the Lyman-$\alpha$ bound
is expected to be insensitive to scalar DM with small fractions
(as discussed in Section~\ref{sec:lyman}),
the bound obtained using the estimator Eqs.~\eqref{eq:deltaA}
and~\eqref{eq:A} improves towards smaller mass
even in the regime $m \lesssim 10^{-22}$ eV.
Nonetheless, the area criterion could be considered as a
simple, fast and intuitive method for performing preliminary tests on
any DM scenario with Lyman-$\alpha$ forest data. Thereby, we leave for
future work a careful study in order to figure out a more accurate
estimator (e.g., by introducing a weight function inside
Eq.~\eqref{eq:A}). 


\end{document}